\documentclass[12pt]{article}
\usepackage[dvips]{graphicx}

\def\lromn#1{\uppercase\expandafter{\romannumeral#1}}
\def\ds{\displaystyle}

\newcommand{\lsim}{ \mathop{}_{\textstyle \sim}^{\textstyle <} }

\begin{document}
  
\begin{titlepage}

\begin{center}

\hfill KEK-TH-1072 \\
\hfill \today

\vspace{1cm}
{\large\bf Cosmic Positron Signature from Dark Matter \\
in the Littlest Higgs Model
with T-parity}
\vspace{1.5cm}

{\bf Masaki Asano}$^{(a,b)}$
\footnote{E-mail: masano@post.kek.jp},
{\bf Shigeki Matsumoto}$^{(a)}$
\footnote{E-mail: smatsu@post.kek.jp},
{\bf Nobuchika Okada}$^{(a,b)}$
\footnote{E-mail: okadan@post.kek.jp},
\\
and
{\bf Yasuhiro Okada}$^{(a,b)}$
\footnote{E-mail: yasuhiro.okada@kek.jp}
\vskip 0.2in
{\it
$^{(a)}${Theory Group, KEK, Oho 1-1 Tsukuba, 305-0801, Japan} \\
$^{(b)}${The Graduate University for Advanced Studies (Sokendai),
\\
Oho 1-1 Tsukuba, 305-0801, Japan}\\
}

\vskip 1in

\abstract{We calculate the flux of cosmic positrons from the dark matter
annihilation in the littlest Higgs model with T-parity. The dark matter
annihilates mainly into weak gauge bosons in the halo, and high energy
positrons are produced through leptonic and hadronic decays of the bosons.
We investigate a possibility to detect the positron signal in upcoming
experiments such as PAMELA and AMS-02. We found that the dark matter signal
can be distinguished from the background in the PAMELA experiment when the
dark matter  mass is less than 120 GeV and the signal flux is enhanced due
to a small scale clustering of dark matter. Furthermore, the signal from the
dark matter annihilation can be detected in the AMS-02 experiment, even if such
enhancement does not exist. We also discuss the invisible width of the Higgs
boson in this model.}

\end{center}
\end{titlepage}
\setcounter{footnote}{0}

\vspace{1.0cm}
\lromn 1 \hspace{0.2cm} {\bf Introduction}
\vspace{0.5cm}

The hierarchy problem in the Standard Model (SM) is expected to give a clue
to explore the physics beyond the SM. This problem is essentially related to
quadratically divergent corrections to the Higgs boson mass, and we need a
mechanism to avoid the divergences. To solve the problem, many scenarios have
been proposed so far, for example supersymmetry, in which the divergences are
completely removed. Other examples are scenarios with a low energy cutoff
scale around a TeV such as Techni-color and TeV scale extra-dimension.

The latter scenarios are, however, constrained by the electroweak precision
measurements. From the analysis of higher dimensional operators at the cutoff
scale, it has been found that the scale should be larger than roughly 5 TeV
\cite{Barbieri}. For such high energy cutoff, the hierarchy problem 
appears again: We still need the fine-tuning of a few percent level in the
Higgs mass term in order to obtain the 100-200 GeV Higgs boson mass. This
problem is called the little hierarchy problem.

Recently the little Higgs model
\cite{Arkani-Hamed:2001nc,Arkani-Hamed:2002qy} has been
proposed for solving the little hierarchy problem. In this scenario, the
Higgs boson is regarded as a pseudo Nambu-Goldstone boson. New particles such
as heavy gauge bosons and a top-partner are introduced, and all quadratic
divergences to the Higgs mass term completely vanish at one-loop level due to
these particles' contributions. Thus, the fine-tuning of the Higgs boson mass
is avoided even if the cutoff scale is around 10 TeV.

The original little Higgs model is still strongly constrained by the
electroweak precision measurements \cite{difficulty}.
This is mainly due to the contributions to electroweak
observables from new heavy gauge bosons, because their masses
are much smaller than the cutoff scale. In particular, direct couplings among
a new heavy gauge boson and SM particles give sizable contributions to the
observables. As a result, masses of new particles have to be raised, and
the fine-tuning of the Higgs boson mass is reintroduced.

To resolve the problem, the implementation of the $Z_2$ symmetry called
T-parity to the model has been proposed \cite{Cheng:2003ju}-\cite{Low:2004xc}.
Under the parity, the new particles are assigned to be $-$ charge (T-odd),
while the SM particles have $+$ charge (T-even). Thanks to the symmetry,
dangerous interactions stated above are prohibited, and the masses of new
particles can be lighter.

Due to the T-parity, the lightest T-odd particle becomes stable and a good
candidate of the dark matter. This is an interesting feature of the model,
because the existence of the dark matter is now established by recent
cosmological observations \cite{Spergel:2003cb}. Since the lightest T-odd
particle is electrically and color neutral, and has a mass of
${\cal O}$(100) GeV \cite{Cheng:2003ju} in many little Higgs models with
T-parity, these models provide a WIMP (weakly interacting massive particle)
dark matter \cite{Jungman:1995df}, and are able to account for the large
scale structure of the present universe \cite{Primack:2002th}.

In this paper, we study the dark matter phenomenology in the littlest Higgs
model with T-parity \cite{Cheng:2004yc,Low:2004xc,Hubisz:2004ft}.
The relic abundance of the dark matter in the thermal relic scenario has
already been evaluated \cite{Hubisz:2004ft}, and it has been found that
the mass of the dark matter consistent with the WMAP observation
\cite{Spergel:2003cb} is around a few hundred GeV. 
In this paper, we focus on the indirect detection of this dark matter using
the cosmic positrons. The dark matter in the halo associated with our galaxy
frequently annihilates and produces high energy particles, for example,
positrons \cite{positrons1}-\cite{Hooper:2004bq}, anti-protons
\cite{antiprotons}, etc. Then high energy positron excess in the cosmic ray
provides an opportunity to search for the dark matter signal.

In the littlest Higgs model with T-parity, the dark matter candidate is a
heavy photon, which annihilates mainly into weak gauge bosons
\cite{Hubisz:2004ft}. Positrons are produced through decays of the bosons.
Since the dark matter annihilation occurs in the s-wave and weak gauge bosons
can produce high energy positrons through leptonic decays, the resultant
positron flux is large and its spectrum becomes harder than that of background
positrons originating in a secondary production from cosmic protons. This
feature is quite different from the spectrum of a bino-like neutralino dark
matter in supersymmetric models, which is expected to be much softer.
In this paper we calculate the positron flux from the dark matter annihilation
in the model, and estimate the possibility to detect these positrons in future
experiments such as PAMELA \cite{pamela} and AMS-02 \cite{Barao:2004ik}.

This paper is organized as follows. In the next section, we briefly review
the littlest Higgs model with T-parity, in particular, focusing on the mass
spectrum in the gauge-Higgs sector and interactions relevant to the
calculation of the dark matter annihilation. We also present the thermal
relic abundance of the dark matter. Calculation of the positron spectrum
from the dark matter annihilation in the halo is performed in Sec.{\lromn 3}
using a diffusion model. Results of the positron flux are shown in
Sec.{\lromn 4}. We also present the $\chi^2$-analysis in order to investigate
a possibility to detect the positron signal in future experiments.
Sec.{\lromn 5} is devoted to summary and discussions including the Higgs decay
into the dark matter. In Appendix, we consider constraints on the model from
electroweak precision measurements, and show that the entire region restricted
by the WMAP observation can be consistent with the measurements.

\vspace{1.0cm}
\lromn 2 \hspace{0.2cm} {\bf Dark Matter in Littlest Higgs Model with
T-parity}
\vspace{0.5cm}

We briefly review the littlest Higgs model with T-parity. In particular, we
focus on the mass spectrum in the gauge-Higgs sector and interactions
relevant to the dark matter in the model. We also calculate the relic
abundance of the dark matter and present a parameter region of the model
consistent with the WMAP observation \cite{Hubisz:2004ft}.
For the general review of little Higgs models and some phenomenological
aspects, see Refs\cite{littlest_review,littlest}.

\vspace{0.5cm}
\underline{\bf Littlest Higgs Model with T-parity}
\vspace{0.5cm}

The littlest Higgs model \cite{Arkani-Hamed:2002qy} is based on a non-linear
sigma model describing an SU(5)/SO(5) symmetry breaking. The non-linear sigma
field $\Sigma$ is given as
\begin{eqnarray}
 \Sigma = e^{2i\Pi/f}\Sigma_0~,
\end{eqnarray}
where $f$ is the vacuum expectation value associated with the symmetry
breaking. The Nambu-Goldstone boson matrix, $\Pi$, and the direction of the
symmetry breaking in the non-linear field, $\Sigma_0$, are written as
\begin{eqnarray}
 \Pi
 =
 \frac{1}{\sqrt{2}}
 \left(
  \begin{array}{ccc}
   0 & H & \sqrt{2}\Phi \\
   H^\dagger &  0 & H^T \\
   \sqrt{2}\Phi^\dagger  & H^* & 0 \\
  \end{array}
 \right)~,
 \qquad
 \Sigma_0
 =
 \left(
  \begin{array}{ccc}
   0 & 0 & {\bf 1} \\
   0 & 1 & 0 \\
   {\bf 1} & 0 & 0 \\
  \end{array}
 \right)~.
 \label{pNG matrix}
\end{eqnarray}
An [SU(2)$\times$U(1)]$^2$ subgroup in the global symmetry SU(5) is gauged,
which is broken down to the diagonal subgroup identified with the SM gauge
group (SU(2)$_L\times$U(1)$_Y$). Due to the presence of the gauge
interactions (and Yukawa interactions if we introduce), the global symmetry
SU(5) is not exact, and the particles in the $\Pi$ field become pseudo
Nambu-Goldstone bosons.

Fourteen ($= 24 - 10$) Nambu-Goldstone bosons are decomposed into
representations under the electroweak gauge group as
${\bf 1}_0\oplus {\bf 3}_0 \oplus {\bf 2}_{\pm 1/2} \oplus {\bf 3}_{\pm 1}$.
The first two representations are real, and become longitudinal components
of gauge bosons
when the [SU(2)$\times$U(1)]$^2$ is broken down to the SM gauge group. The
representations ${\bf 2}_{\pm 1/2}$ and ${\bf 3}_{\pm 1}$ are a complex
doublet identified with the SM Higgs field ($H$ in Eq.(\ref{pNG matrix})) and
a complex triplet Higgs field ($\Phi$ in Eq.(\ref{pNG matrix})), respectively.

The kinetic term for the $\Sigma$ field is given as
\begin{eqnarray}
 {\cal L}_{\Sigma}
 =
 \frac{f^2}{8}{\rm Tr}
 \left[
  D_\mu\Sigma\left(D^\mu\Sigma\right)^\dagger
 \right]~,
 \label{Kinetic L}
\end{eqnarray}
where
\begin{eqnarray}
 D_\mu\Sigma
 =
 \partial_\mu\Sigma
 -
 i
 \sum_{j = 1}^2
 \left[
  g_j W^a_j(Q_j^a\Sigma + \Sigma Q^{a T}_j)
  +
  g'_j B_j(Y_j\Sigma + \Sigma Y_j)
 \right]~.
\end{eqnarray}
Here, $W^a_j(B_j)$ are the SU(2)$_j$(U(1)$_j$) gauge fields and $g_j(g'_j)$
are corresponding gauge coupling constants. The generators of the gauge
symmetries $Q_j$ and $Y_j$ are
\begin{eqnarray}
 &&
 Q_1^a
 =
 ~~~
 \frac{1}{2}
 \left(
  \begin{array}{ccc}
   \sigma^a & 0 & 0
   \\
   0 & 0 & 0
   \\
   0 & 0 & 0
  \end{array}
 \right)~,
 \qquad
 Y_1
 =
 {\rm diag}(3,3,-2,-2,-2)/10~,
 \nonumber \\
 &&
 Q_2^a
 =
 -\frac{1}{2}
 \left(
  \begin{array}{ccc}
   0 & 0 & 0
   \\
   0 & 0 & 0
   \\
   0 & 0 & \sigma^{a*}
  \end{array}
 \right)~,
 \qquad
 Y_2
 =
 {\rm diag}(2,2,2,-3,-3)/10~,
\end{eqnarray}
where $\sigma^a$ are the Pauli matrices.

In terms of above fields, the symmetry of the T-parity
\cite{Cheng:2003ju}-\cite{Low:2004xc} is defined as the invariance of the
Lagrangian under the transformation:
\begin{eqnarray}
 W^a_1 \leftrightarrow W^a_2~,
 \qquad
 B_1 \leftrightarrow B_2~,
 \qquad
 \Pi \leftrightarrow -\Omega\Pi\Omega~,
\end{eqnarray}
where $\Omega = {\rm diag}(1,1,-1,1,1)$. As a result of the symmetry, the
gauge coupling $g_1(g'_1)$ must be equal to $g_2(g'_2)$, namely $g_1 = g_2 = 
\sqrt{2}g~(g'_1 = g'_2 = \sqrt{2}g')$, where $g(g')$ is nothing but the
coupling constant of the SM SU(2)$_L$(U(1)$_Y$) gauge symmetry.

Since the Higgs boson is the pseudo Nambu-Goldstone boson, its potential is
generated radiatively \cite{Arkani-Hamed:2002qy,Hubisz:2004ft}
\begin{eqnarray}
 V(H, \Phi)
 =
 \lambda f^2{\rm Tr}\left[\Phi^\dagger\Phi\right]
 -
 \mu^2H^\dagger H
 +
 \frac{\lambda}{4}\left(H^\dagger H\right)^2
 +
 \cdots~.
 \label{Higgs Potential}
\end{eqnarray}
Due to the little Higgs mechanism, quadratic divergent corrections do not
contribute to the Higgs mass $\mu^2$ at 1-loop level, while the corrections
do contribute to the triplet Higgs mass term. Main contributions to $\mu^2$
come from the logarithmic divergent corrections at 1-loop level and quadratic
divergent corrections at 2-loop level. As a result, $\mu^2$ is expected to be
smaller than $f^2$, while the triplet Higgs mass term is proportional to
$f^2$. The quartet coupling $\lambda$ is determined by the 1-loop effective
potential from gauge and top sectors. Since both $\mu$ and $\lambda$ depend
on parameters at the cutoff scale, we treat these as free parameters in this
paper.

We discuss the mass spectrum of gauge and Higgs bosons. This model contains
four kinds of gauge fields $W^a_1$, $W^a_2$, $B_1$ and $B_2$ in the
electroweak gauge sector. The combinations, $W^a = (W^a_1 + W^a_2)/\sqrt{2}$
and $B = (B_1 + B_2)/\sqrt{2}$, correspond to the SM gauge bosons for the
SU(2)$_L$ and U(1)$_Y$ symmetry. The other combinations, $W^a_H = (W^a_1 -
W^a_2)/\sqrt{2}$ and $B_H = (B_1 - B_2)/\sqrt{2}$, are additional gauge
bosons, which acquire the masses of ${\cal O}(f)$ through the SU(5)/SO(5)
symmetry breaking. After electroweak symmetry breaking, neutral components of
$W^a_H$ and $B_H$ are mixed and form mass eigenstates $A_H$ and $Z_H$. The
masses of the heavy bosons are obtained as
\begin{eqnarray}
 m_{Z_H}
 &=&
 \frac{1}{2}
 \left(
  A + C
  +
  \sqrt{(A - C)^2 + 4B^2}
 \right)
 \simeq
 g f~,
 \nonumber \\
 m_{A_H}
 &=&
 \frac{1}{2}
 \left(
  A + C
  -
  \sqrt{(A - C)^2 + 4B^2}
 \right)
 \simeq
 \frac{g'}{\sqrt{5}}f~,
 \label{AH mass}
\end{eqnarray}
where $A = g^2(f^2 - v^2/4)$, $B = g g' v^2/4$ and $C = g^{\prime 2}(f^2/5 -
v^2/4)$. The mixing angle between $W^a_H$ and $B_H$ are given as
\begin{eqnarray}
 \tan\theta_H
 =
 -\frac{2B}{A - C + \sqrt{(A - C)^2 + 4B^2}}
 \simeq
 -\frac{g g' v^2}{4f^2(g^2 - g^{\prime 2}/5)}~,
\end{eqnarray}
which is suppressed by ${\cal O}(v/f)$. In addition to these gauge fields,
we have the triplet Higgs boson $\Phi$ in this model, and its mass is given
by $m_\Phi^2 = \lambda f^2 = 2m_h^2f^2/v^2$, where $m_h$ is the mass of the
SM Higgs boson and $v~(\simeq 246~{\rm GeV})$ is the vacuum expectation value
of the Higgs field. New heavy gauge bosons and the triplet Higgs boson are
T-odd particles, while SM particles are T-even.

The mass spectrum of T-odd particles are determined by two parameters, the
breaking scale ($f$) and the Higgs boson mass ($m_h$). For instance, in the
case of $m_h =$ 120 GeV and $f =$ 700 GeV, $m_{A_H} =$ 100 GeV, $m_{W_H(Z_H)}
=$ 450 GeV and $m_\Phi =$ 500 GeV. As shown in Eq.(\ref{AH mass}), the mass of
the heavy photon is considerably lighter than other T-odd particles due to the
small hypercharge. Thus its stability is guaranteed by the T-parity
conservation and becomes a candidate of a non-baryonic cold dark matter.

In addition to these new particles, top-partners are introduced in this model
in order to cancel the quadratic divergent contribution to the Higgs mass
term from the top quark loop diagrams. Due to the T-parity, three kinds of
partners exist, namely T-even heavy top ($T_+$) which is introduced for the
cancellation, T-partners of heavy top and top quark ($T_-$ and $t_-$). The
mass spectrum of these particles depends not only on $f$ and $m_h$, but also
on other model parameters. Since these particles do not play a significant 
role in the dark matter phenomenology, we do not discuss this sector here.

\vspace{0.5cm}
\underline{\bf Relic Abundance of Dark Matter}
\vspace{0.5cm}

\begin{figure}[t]
\begin{center}
\scalebox{0.8}{\includegraphics*{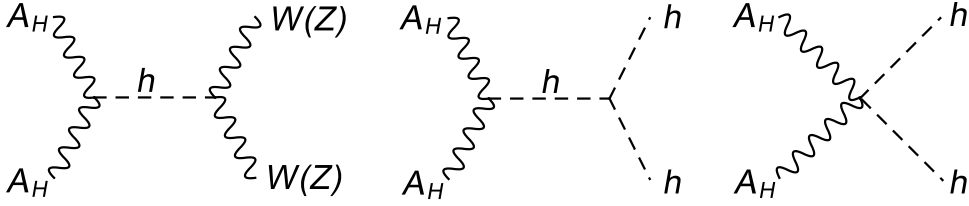}}
\caption{\small Feynman diagrams for the annihilation of the dark matter
         ($A_H$).}
\label{diagrams}
\end{center}
\end{figure}

The dark matter ($A_H$) in the model annihilates mainly into weak gauge
bosons, $W^+W^-, ZZ$ through the diagrams in which the Higgs boson propagates
in the s-channel. The dark matter also annihilates into Higgs bosons if
$m_{A_H} > m_h$. The Feynman diagrams for these processes are shown in
Fig.\ref{diagrams}\footnote{There are also diagrams in which the T-partners
of fermions are exchanged in the t-channel. These contributions are, however,
negligible compared to those in Fig.\ref{diagrams} unless masses of
T-partners are much smaller than 1 TeV. In fact, as discussed in
Ref.\cite{Birkedal:2006fz}, the cross sections for these processes are
suppressed by masses of the T-partners, $m_{f_H}^4$, and small hypercharges,
$\tilde{Y}^4 = (0.1)^4$.}. From Eqs.(\ref{Kinetic L}) and (\ref{Higgs
Potential}), interactions relevant to the annihilation are given as
\begin{eqnarray}
 {\cal L}_{\rm int}
 &=&
 c
 \left(v h + \frac{h^2}{2}\right)A_H^2
 +
 \frac{g^2v}{2}h W^+W^-
 +
 \frac{(g^2 + g^{\prime 2})v}{4}h Z^2
 -
 \frac{m_h^2}{2v}h^3~,
\end{eqnarray}
where $c= -(g\sin\theta_H - g'\cos\theta_H)^2/4$. We have used the unitary
gauge $H = (0, v + h)^T/\sqrt{2}$. From the interactions, the annihilation
cross sections of the dark matter turn out to be
\begin{eqnarray}
 \left.\sigma {\rm v}\right|_{WW}
 &=&
 \frac{1}{96\pi m_{A_H}^2}
 \frac{(g^2v^2c)^2}{(4m_{A_H}^2 - m_h^2)^2 + m^2_h\Gamma_h^2}
 \left(
  4\frac{m_{A_H}^4}{m_W^4}
  -
  4\frac{m_{A_H}^2}{m_W^2}
  +
  3
 \right)
 \sqrt{1 - \frac{m_W^2}{m_{A_H}^2}}~,
 \nonumber \\
 \nonumber \\
 \left.\sigma {\rm v}\right|_{ZZ}
 &=&
 \frac{1}{192\pi m_{A_H}^2}
 \frac{\left[(g^2 + g^{\prime 2})v^2c\right]^2}
 {(4m_{A_H}^2 - m_h^2)^2 + m^2_h\Gamma_h^2}
 \left(
  4\frac{m_{A_H}^4}{m_Z^4}
  -
  4\frac{m_{A_H}^2}{m_Z^2}
  +
  3
 \right)
 \sqrt{1 - \frac{m_Z^2}{m_{A_H}^2}}~,
 \nonumber \\
 \nonumber \\
 \left.\sigma {\rm v}\right|_{hh}
 &=&
 \frac{c^2}{48\pi m_{A_H}^2}
 \left|
  1
  +
  \frac{3m_h^2}{4m_{A_H}^2 - m_h^2 + i m_h\Gamma_h}
 \right|^2
 \sqrt{1 - \frac{m_h^2}{m_{A_H}^2}}~,
 \label{Ann CS}
\end{eqnarray}
where ${\rm v}$ is the relative velocity between incident dark matters, and
$\Gamma_h$ is the width of the SM Higgs boson. We take the non-relativistic
limit (${\rm v}\rightarrow 0$) in the calculation, because the dark matter
is almost at rest at the freeze-out temperature.

The relic abundance of the dark matter is obtained by solving the following
Boltzmann equation \cite{Kolb:1990vq},
\begin{eqnarray}
 \frac{d Y}{d x}
 =
 -\frac{\langle \sigma {\rm v} \rangle}{H x}s
 \left( Y^2 - Y_{\rm eq}^2 \right)~,
 \label{eq:Boltzmann}
\end{eqnarray}
where $Y = n/s$ is the yield of the dark matter defined by the ratio of
the dark matter density ($n$) to the entropy density of the universe ($s =
0.439g_*m_{A_H}^3/x^3$), $g_* = 86.25$ and $x \equiv m_{A_H}/T$ ($T$ is the
temperature of the universe). The Hubble parameter is given by $H =
1.66g_*^{1/2}m_{A_H}^2m_{\rm Pl}/x^2$, where $m_{\rm Pl} = 1.22\times 10^{19}$
GeV is the Planck mass. The yield in the equilibrium $Y_{\rm eq}$ is written
as
\begin{eqnarray}
 Y_{\rm eq}
 =
 \frac{45}{2\pi^4}
 \left(\frac{\pi}{8}\right)^{1/2}
 \frac{3}{g_*}
 x^{3/2} e^{-x}~.
\end{eqnarray}
Since the dark matter annihilates into the SM particles in the s-wave at the
non-relativistic limit, the thermal averaged annihilation cross section
($\langle\sigma {\rm v}\rangle$) is simply given by
\begin{eqnarray}
 \langle\sigma {\rm v}\rangle
 =
 \left.\sigma {\rm v}\right|_{WW}
 +
 \left.\sigma {\rm v}\right|_{ZZ}
 +
 \left.\sigma {\rm v}\right|_{hh}~,
\end{eqnarray}
After solving the Boltzmann equation, we obtain the present abundance of
dark matter ($Y_\infty$). It is useful to express the relic density in terms
of the ratio of the dark matter density to the critical density ($\Omega h^2
= m_{A_H} n h^2/\rho_c = m_{A_H}s_0Y_\infty h^2/\rho_c$), where $\rho_c = 1.1
\times 10^{-5} h^2$ cm$^{-3}$, $h = 0.71^{+0.04}_{-0.03}$ and $s_0 = 2900$
cm$^{-3}$. With a good accuracy, the solution of Eq.(\ref{eq:Boltzmann}) is
approximately given as
\begin{eqnarray}
 \Omega h^2
 &=&
 \frac{1.07\times 10^9 x_f{\rm GeV}^{-1}}
 {\sqrt{g_*}m_{\rm PL}\langle\sigma {\rm v}\rangle}~,
\end{eqnarray}
where $x_f = m_{A_H}/T_f$ is the freeze-out temperature for the dark matter
and given as $x_f = \ln(X) - 0.5\ln(\ln(X))$ with $X =
0.038\cdot (3/g_*^{1/2})m_{\rm PL}m_{A_H}\langle\sigma {\rm v}\rangle$.
Typically  $x_f$ takes a value, $x_f\simeq 23$.

\begin{figure}[t]
\begin{center}
\scalebox{0.6}{\includegraphics*{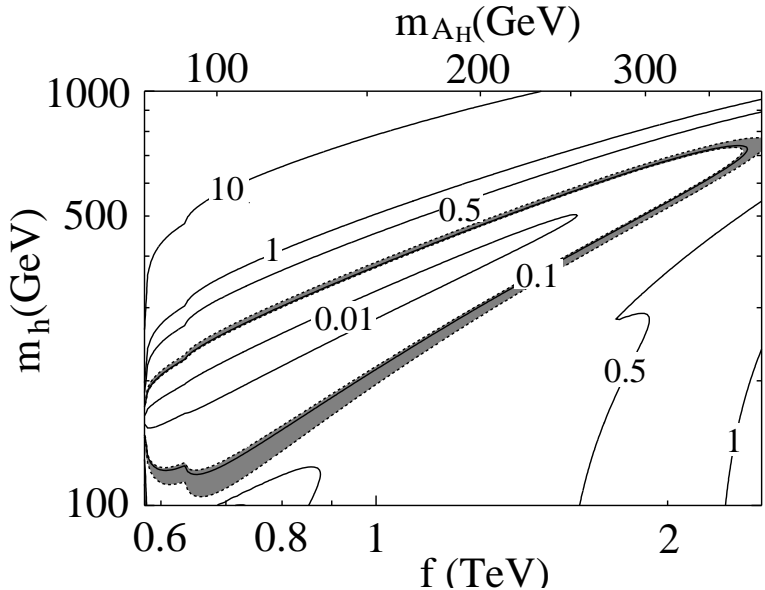}}
\caption{\small Contour plot of the thermal relic abundance for the dark
         matter ($\Omega h^2$) in the $(f,m_h)$-plane. The shaded thin area
         is the allowed region from the WMAP observation at 2$\sigma$ level,
         $0.094 < \Omega h^2 < 0.129$.}
\label{fig:abundance}
\end{center}
\end{figure}

The relic abundance of the dark matter in the thermal scenario is depicted in
Fig.\ref{fig:abundance} as a contour map in the $(f,m_h)$-plane. The shaded
thin area is the allowed region for the WMAP observation at 2$\sigma$ level,
$0.094 < \Omega h^2 < 0.129$ \cite{Spergel:2003cb}. The result obtained here
is consistent with the previous calculation in Ref.\cite{Hubisz:2004ft}.
As shown in the figure, the breaking scale $f$ is constrained to be less than
about 2 TeV.

The littlest Higgs model with T-parity is constrained from the electroweak
precision measurements. New physics contributions to the electroweak
observables come from radiative corrections, because there is no tree-level
effect due to the T-parity. As a result, the constraint to the model becomes
weaker than that for the model without T-parity. The detailed analysis has been
performed in Ref.\cite{Hubisz:2005tx}. We have repeated this analysis and
present the result in Appendix. It is shown that the entire region is
consistent with the electroweak precision measurements by choosing parameters
in the top sector.

\vspace{1.0cm}
\lromn 3 \hspace{0.2cm} {\bf Propagation of Positron from Dark Matter
Annihilation in Galaxy}
\vspace{0.5cm}

In the present universe, a dark matter makes up a halo associated with a
galaxy. The distribution of the dark matter in the halo is given from a halo
mass profile $\rho(\vec{r})$ through the equation $n(\vec{r}) =
\rho(\vec{r})/m$. The profile is determined by observations of the rotational
velocity of the galaxy and the motions of the dwarf galaxies with help of
N-body simulations. Several models for the profile have been proposed
\cite{profile}. For our galaxy, we use the isothermal halo model in this
paper, which is
given as
\begin{equation}
 \rho(\vec{r})
 =
 \rho_0\frac{1 + r_0^2/r_c^2}{1 + r^2/r_c^2}~~
 ({\rm GeV/cm^3})~,
\label{isothermal}
\end{equation}
where $r = |\vec{r}|$ is the distance from the galactic center, $\rho_0
\simeq$ 0.43~GeV/cm$^3$ is the local halo density in the vicinity of the
solar system, $r_c \simeq$ 2.8 kpc is the core radius of the galaxy, and $r_0
\simeq$ 8.5 kpc is the distance between the galactic center and the solar
system. 

In an indirect detection of dark matter, high energy particles from the dark
matter annihilation are expected to be observed in the cosmic ray. Several
kinds of annihilation products are produced, among which we focus on
positrons \cite{positrons1}. We calculate the expected flux of the positrons
at the earth from the annihilation. In evaluation of the flux, we need to
take into account the propagation of positrons through the galaxy. We also
address the background positrons originated from the secondary production of
the cosmic ray.

\vspace{0.5cm}
\underline{\bf Production Rate of Positrons from Dark Matter Annihilation}
\vspace{0.5cm}

The dark matter annihilates mainly into weak gauge bosons and the Higgs
bosons as discussed in the previous section. Positrons are produced through
leptonic and hadronic cascade decays of these bosons. For $W$ bosons, these
processes are $W^+ \rightarrow e^+\nu$, $W^+ \rightarrow \mu^+\nu \rightarrow
e^+\nu \bar{\nu}\nu$ or $W^\pm \rightarrow {\rm hadrons} \rightarrow \pi^\pm
\rightarrow \mu^\pm \rightarrow e^\pm$. Decay branching ratio for the Higgs
boson depends on its mass. When $m_h > 160$ GeV, the Higgs boson mainly
decays into weak gauge bosons, and positrons are produced by decays of weak
bosons.

Using the annihilation cross sections in Eq.(\ref{Ann CS}), the production
rate of positrons from the annihilation is given as
\begin{equation}
 Q(E,\vec{r})
 =
 \frac{1}{2} n^2(\vec{r})
 \sum_{f = WW, ZZ, hh}
 \left(\sigma_f v\right)
 \left(\frac{d N_{e^+}}{d E}\right)_f~,
\end{equation}
where $E$ is the energy of a positron and the coefficient $1/2$ comes from
the pair annihilation of the identical particles. The fragmentation function
$(d N_{e^+}/d E)_f$ represents the number of positrons with energy $E$
produced from the final state $f$. The cascade processes for the positron
production discussed above are encoded into these functions.

The fragmentation functions are evaluated by a Monte-Carlo simulation such
as the HERWIG code \cite{Corcella:2000bw}. These functions for the weak
gauge bosons can be parameterized by a single variable $x = E/m$. The
fitting functions have been constructed in Ref.\cite{Hisano:2005ec} to
reproduce results of the simulation. We use these functions in this paper.
The fragmentation function for the Higgs boson is also obtained by assuming
the dominance of gauge boson decay modes.

Although we use the isothermal model in Eq.(\ref{isothermal}), the high
energy positron flux does not strongly depend on the choice of dark matter
halo models. Main difference among proposed models appears in the region
around the galactic center, and positrons produced around this region can not
reach the earth without the significant energy loss.

Recently, the effect of inhomogeneity in the local dark matter distribution
on the positron flux is discussed based on the N-body simulations. It is
shown that the positron flux from the dark matter annihilation is enhanced
if there are clumps of the dark matter in the vicinity of the solar system
\cite{Boost Factor}. The effect is parameterized as a boost factor ($BF$),
which is defined by the ratio of the signal fluxes with inhomogeneity and
without inhomogeneity,
\begin{eqnarray}
 BF
 =
 \frac{V\ds\int_V d^3x~\rho^2}
 {\left(\ds\int_V d^3x~\rho\right)^2}~,
\end{eqnarray}
where the region of the integration is taken to be $V \sim$ (a few kpc)$^3$
around the solar system. The boost factor is larger than 1, and equal to 1
only if the density $\rho$ is a constant. The value of the factor is expected
to be in the range of 2 to 5 based on a hierarchical clustering scenario in
the inflationary universe \cite{Berezinsky:2003vn}.

\vspace{0.5cm}
\underline{\bf Positron Propagation in Galaxy}
\vspace{0.5cm}

Once positrons are produced in the dark matter annihilation, they travel in
our galaxy under the influence of a tangled magnetic field. Since the
typical strength of the magnetic field is a micro Gauss, the gyro-radius of
the positron is much less than the galactic radius. Thus, the propagation can
be treated as a random walk.

We use a diffusion model for the propagation of positrons,  in which the
random walk is described by the following diffusion equation
\cite{posbaltz,Hisano:2005ec},
\begin{equation}
 \frac{\partial}{\partial t}f_{e^+}(E,\vec{r})
 =
 K(E)\nabla^2 f_{e^+}(E,\vec{r})
 +
 \frac{\partial}{\partial E}
 \left[b(E)f_{e^+}(E,\vec{r})\right]
 +
 Q(E,\vec{r})~,
 \label{eq:diffusion}
\end{equation}
where $f_{e^+}(E,\vec{r})$ is the number density of positrons per unit energy,
$E$ is the energy of positron, $K(E)$ is the diffusion constant, $b(E)$ is the
energy loss rate, and $Q(E,\vec{r})$ is the source (positron injection) term
discussed in the previous subsection. The flux of positrons with high energy
($E \gg m_e$) in the vicinity of the solar system is given from
$f_{e^+}(E,\vec{r})$ as
\begin{equation}
 \Phi_{e^+}(E)
 =
 BF\frac{1}{4\pi}f_{e^+}(E,\vec{r}_\odot)~,
 \label{flux}
\end{equation}
where $\vec{r}_\odot$ represents the coordinate of the solar system.

There are two parameters in Eq.(\ref{eq:diffusion}). One is the diffusion
constant $K(E)$ characterizing the tangled magnetic field of the galaxy. This
parameter is evaluated by comparing the observations of the Boron to Carbon
ratio in the cosmic ray with the result of simulations \cite{BtoC}. The
parameter $b(E)$ stands for the energy loss rate of positrons due to the
inverse Compton scattering with cosmic microwave radiation (and infrared
photons from stars) and synchrotron radiation with the magnetic field
during the propagation in the galaxy \cite{Longair:1994wu}. This parameter
is, therefore, determined by the photon density, the strength of the magnetic
field and the Thomson scattering cross section. For both parameters, we use
values adopted in Ref.\cite{posbaltz}
\begin{eqnarray}
 K(E)
 &=&
 3.3\times 10^{27}
 \left[3^{0.6}+ (E / 1~{\rm GeV} )^{0.6}\right]~~
 ({\rm cm^2s^{-1}})~,
 \nonumber \\
 b(E)
 &=&
 10^{-16} (E / 1~{\rm GeV})^2~~
 ({\rm  GeV s^{-1}})~.
\end{eqnarray}

The positrons from the dark matter annihilation are expected to be in the
equilibrium in our galaxy, hence the number density
$f_{e^+}(E,\vec{r})$ is obtained by solving Eq.~(\ref{eq:diffusion}) with the
steady state condition $\partial f_{e^+}/\partial t = 0$. Furthermore, we
impose the free escape boundary condition, namely the positron density drops
to zero on the surface of the diffusion zone. This means that the positrons
coming from the outside of the zone are negligible, while the positrons
produced inside the zone contribute to the flux around the solar system due
to the trapping by the tangled magnetic field \cite{Barrau:2001ev}. The
diffusion zone is usually assumed to be a cylinder with the half-height ($L$)
and radius ($R$), which are set to be $L = 4$ kpc and $R = 20$ kpc in this
paper.

The high energy positron flux at the earth does not strongly depend on the
choice of the diffusion zone, because the positrons we observe are produced
within a few kpc around the solar system. In fact, the distance in which
positrons travel without significant energy loss is estimated as
$r\sim\sqrt{E K(E)/b(E)}\sim 0.74\times(E/{\rm 100~GeV})^{-0.2}$ kpc.
Positrons far from the earth lose their energies during the propagation, and
consequently they only contribute to the low-energy part of the flux.

The flux obtained from Eq.(\ref{flux}) does not correspond exactly to the one
observed on the top of atmosphere. The flux is modified due to interaction
with the solar wind and the magneto-sphere \cite{solarmod1}. However, the
modulation effect is not important when the energy of a positron is above 10
GeV. Furthermore, the effect is highly suppressed in the positron fraction,
which is defined by a ratio of the positron flux to the sum of positron and
electron fluxes, {\it i.e.} $\Phi_{e^+}/(\Phi_{e^+} + \Phi_{e^-})$.

\vspace{0.5cm}
\underline{\bf Background Positrons}
\vspace{0.5cm}

In the detection of the positrons from the dark matter annihilation, the main
background comes from high energy positrons in the cosmic ray. High energy
positrons are produced as secondary particles in the collision between
hydrogen and helium in interstellar medium and primary particles in cosmic
ray accelerated by the shock wave in supernovas. The flux of these positrons
are obtained by simulations, in which a diffusion model is also used.
The result of simulations agrees with measurements of the low-energy positron
flux in the cosmic ray \cite{Moskalenko:1997gh}. The fitting
functions for high energy positrons, primary electrons, and secondary
electrons have been constructed \cite{posbaltz}, 
\begin{eqnarray}
 \Phi^{\rm (prim)}_{e^-}(E)
 &=&
 \frac{0.16 E^{-1.1}}
      {1+11 E^{0.9}+3.2 E^{2.15}}~~
      ({\rm GeV^{-1}cm^{-2}s^{-1}sr^{-1}})~,
 \nonumber \\
 \nonumber \\
 \Phi^{\rm (sec)}_{e^{-}}(E)
 &=&
 \frac{0.70 E^{0.7}}
 {1+110 E^{1.5}+600 E^{2.9}+580 E^{4.2}}~~
 ({\rm GeV^{-1}cm^{-2}s^{-1}sr^{-1}})~,
 \nonumber \\
 \nonumber \\
 \Phi^{\rm (sec)}_{e^+}(E)
 &=&
 \frac{4.5 E^{0.7}}
 {1+650E^{2.3}+1500 E^{4.2}}~~
 ({\rm GeV^{-1}cm^{-2}s^{-1}sr^{-1}})~,
 \label{background}
\end{eqnarray}
where $E$ is in unit of GeV. The first one, $\Phi^{\rm (prim)}_{e^-}$, is the
flux of the primary electrons, while the second and third ones, $\Phi^{\rm
(sec)}_{ e^-}$ and $\Phi^{\rm (sec)}_{ e^+}$, are the secondary electron and
positron fluxes, respectively.

\vspace{1.0cm}
\lromn 4 \hspace{0.2cm} {\bf Positron Signal from Dark Matter Annihilation
in Halo}
\vspace{0.5cm}

\begin{figure}[t]
\begin{center}
\scalebox{0.6}{\includegraphics*{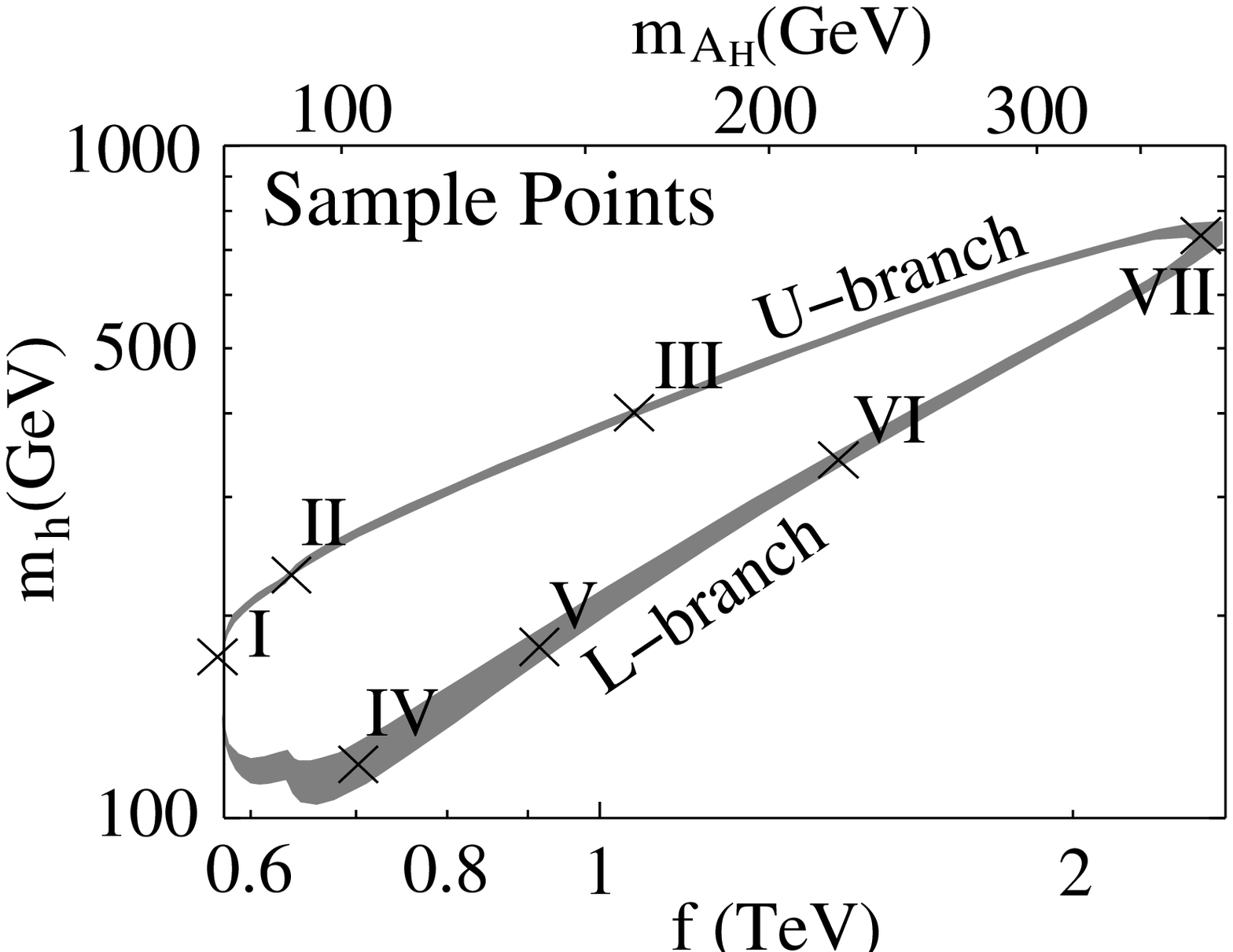}}
\qquad~~~ \\
~\\
\begin{tabular}{|c||c|c|c|c|c|c|c|}
 \hline
         (GeV unit) &    I &   II &  III &  IV &   V &   VI & VII \\
 \hline
                $f$ &  577 &  637 & 1050 & 702 & 916 & 1418 & 2470 \\
 \hline
              $m_h$ &  170 &  230 &  400 & 120 & 180 &  340 & 740 \\
 \hline
          $m_{A_H}$ & 80.3 & 91.2 &  162 & 103 & 139 &  222 & 392 \\
 \hline
 $m_{W_H}(m_{Z_H})$ &  367 &  406 &  678 & 450 & 590 &  919 & 1600 \\
 \hline
           $m_\Phi$ &  564 &  842 & 2410 & 484 & 948 & 2770 & 10500 \\
 \hline
\end{tabular}
\caption{\small Sample-points for depicting the positron fraction from the
         dark matter annihilation. The shaded thin area is the allowed
         region from the WMAP observation at 2$\sigma$ level. In the table
         below the figure, the details about model parameters in each point
         are shown.}
\label{fig:sample_points}
\end{center}
\end{figure}

We are now in position to discuss the positron signal from the dark matter
annihilation. The signal positron flux is evaluated in Eq.(\ref{flux}), while
expected positron and electron background are given in Eqs.(\ref{background}).
In order to show how the dark matter annihilation can modify the positron 
energy spectrum in the cosmic ray, we have chosen seven sample-points (I to
VII) in the parameter space of the little Higgs model with T-parity as in
Fig.\ref{fig:sample_points}. Parameters $f$ and $m_h$ and masses of various
particles are listed in the table below the figure. All points satisfy the
WMAP condition, namely the present dark matter abundance is explained by the
thermal relic scenario. As seen in the figure, there are two branches: the
upper branch (U-branch) and lower branch (L-branch). In the U-branch, the
Higgs boson mass is larger than twice the dark matter mass, $m_h > 2m_{A_H}$,
while $m_h < 2m_{A_H}$ in the L-branch.

\begin{figure}[t]
 \begin{center}
  \includegraphics[width=7.2cm]{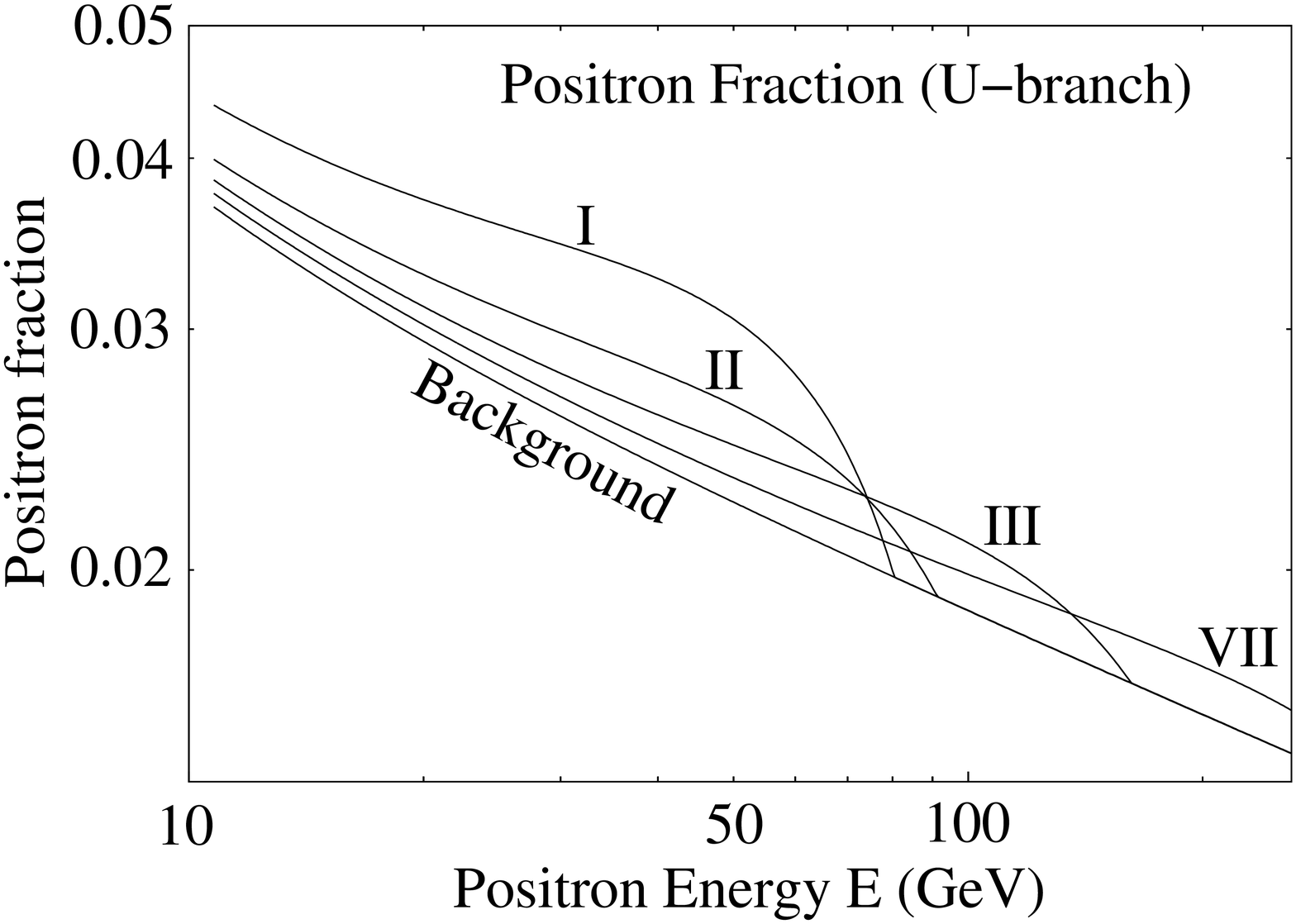}
  ~
  \includegraphics[width=7.2cm]{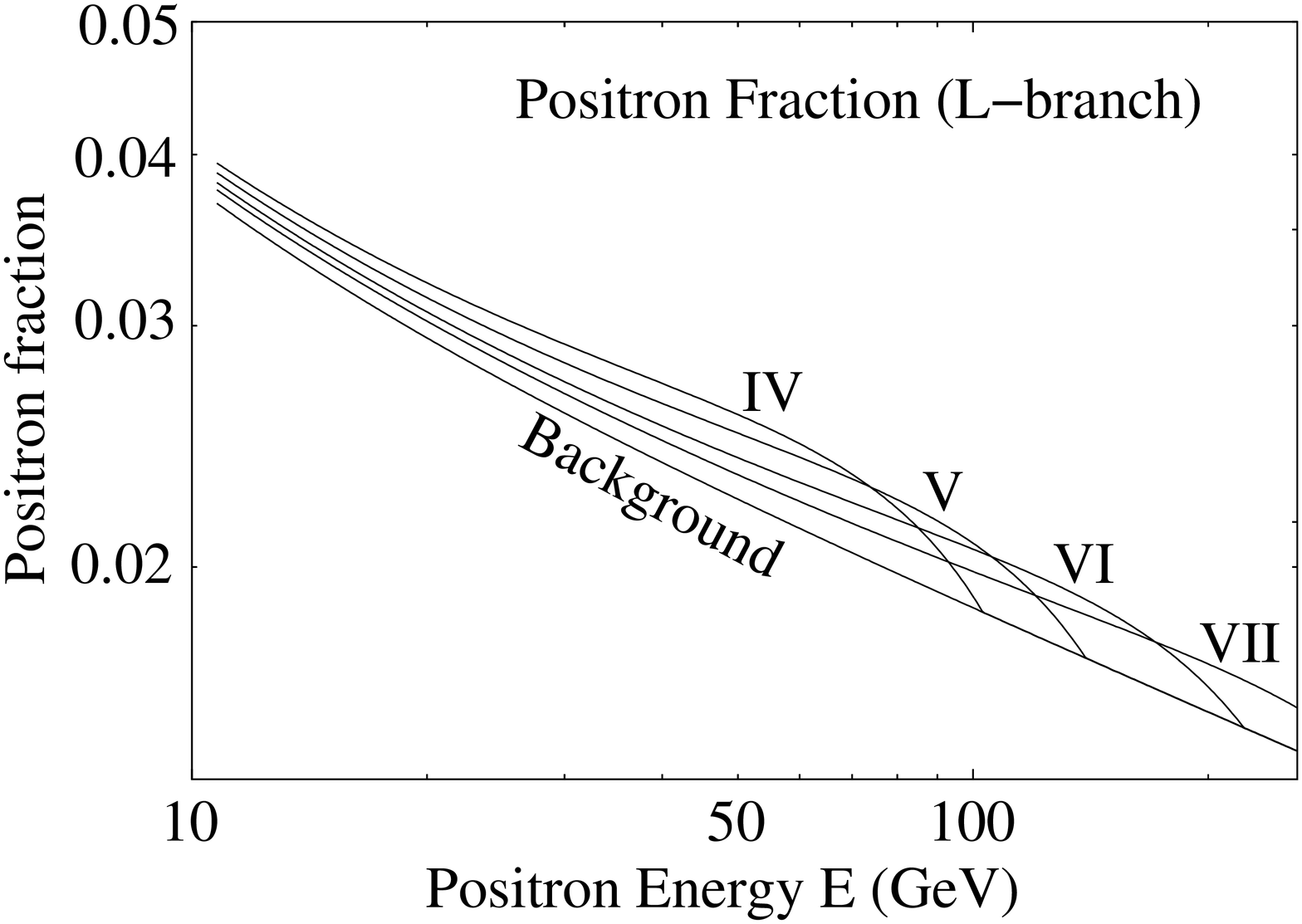}
  \caption{\small The positron fraction as a function of positron energy $E$.
   For comparison, the expected background fraction is also shown in these
   figures. In the left figure, the fraction in the U-branch
   (I to III and VII) are depicted, while those in the L-branch
   (IV to VI and VII) are in the right figure. In both figures, the boost
   factor $BF = 5$ is used.}
  \label{positron fraction}
 \end{center}
\end{figure}

In Fig.\ref{positron fraction}, the positron fraction,
$\Phi_{e^+}/(\Phi_{e^+} + \Phi_{e^-})$, is shown as a function of positron
energy. In the left figure, the results for the points in the U-branch (I to
III) are depicted, while those in the L-branch (IV to VI) are in the right
figure. The point VII can be regarded as a sample on both U- and L-branches,
and its result are shown in both figures. We used the boost factor $BF = 5$.
The expected background positron fraction is also shown for comparison.

The upcoming experiments such as PAMELA and AMS-02 have good sensitivities in
a broad range for a positron energy 10 GeV $\leq E \leq$ 270 GeV. In order to
discuss the possibility for detection of the dark matter signal in the future
experiments, we perform the $\chi^2$-analysis developed in
Ref.\cite{Hooper:2004bq}. For this purpose, we need to know the expected
signal and background events in future experiments for each parameter point of
the model ($f$, $m_h$). The $\chi^2$ is defined as
\begin{eqnarray}
 \chi^2
 =
 \sum_i\frac{(N^{({\rm Obs})}_i - N^{({\rm BG})}_i)^2}{N^{({\rm Obs})}_i}~,
\end{eqnarray}
where the sum is taken over energy bins, $N^{({\rm Obs})}_i$ is the number of
positron events observed in the i-th bin and $N^{({\rm BG})}_i$ is the
number of events expected from the background contribution in the bin.
Following the Ref.\cite{Hooper:2004bq}, we chose 22 bins in the range between
10 GeV $< E <$ 270 GeV,
\begin{eqnarray}
 &&
 \Delta[\log_{10}(E/1{\rm GeV})] = 0.06~,
 \qquad~
 (E \le 40 {\rm GeV})~,
 \nonumber \\
 &&
 \Delta[\log_{10}(E/1{\rm GeV})] = 0.066~,
 \qquad
 (E > 40 {\rm GeV})~.
\end{eqnarray}
In our analysis, we use the acceptance of PAMELA and AMS-02 to be
20.5cm$^2$sr and 450cm$^2$sr, respectively, assuming three years of
data-taking. 

\begin{figure}[t]
 \begin{center}
  \includegraphics[width=7.3cm]{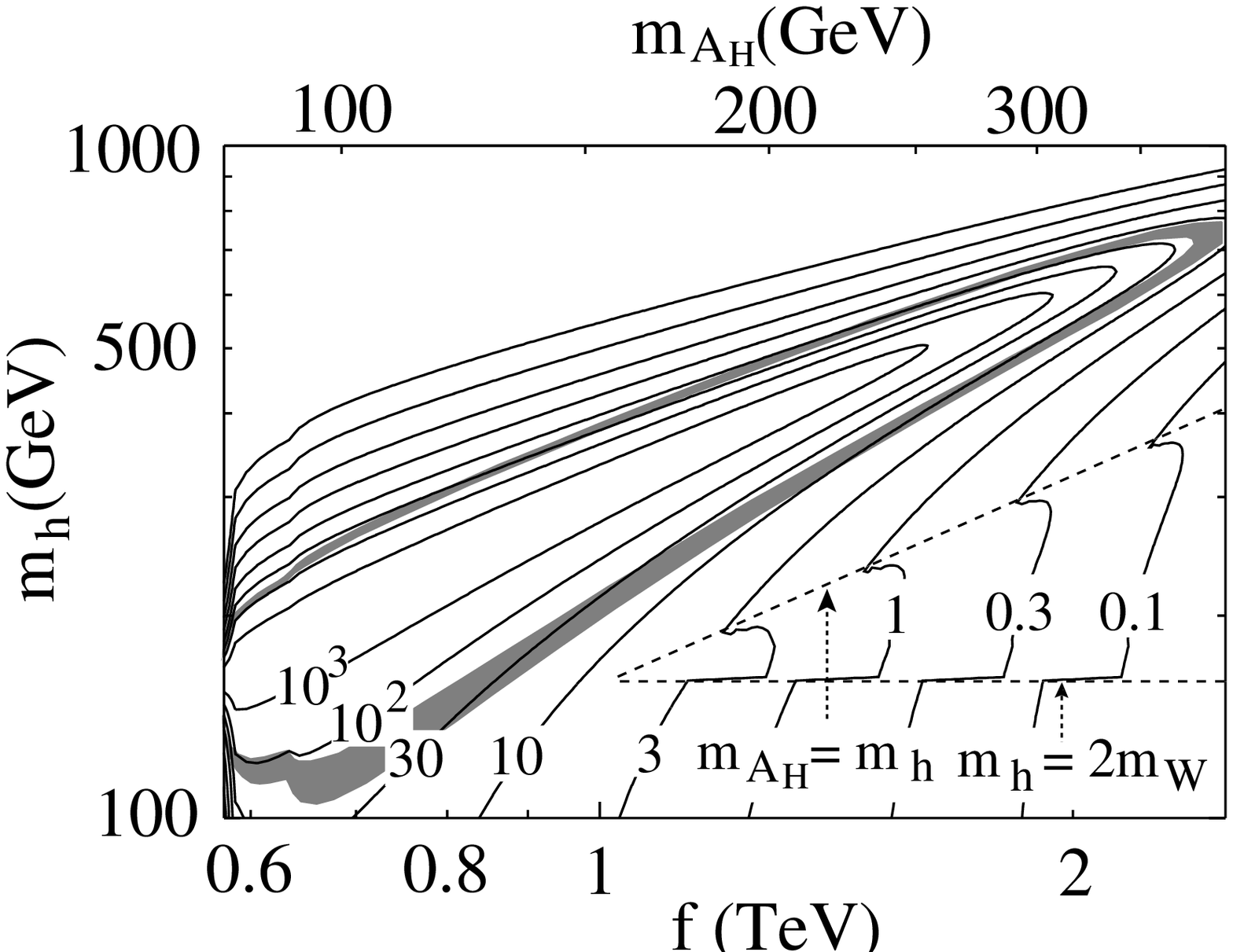}
  ~
  \includegraphics[width=7.3cm]{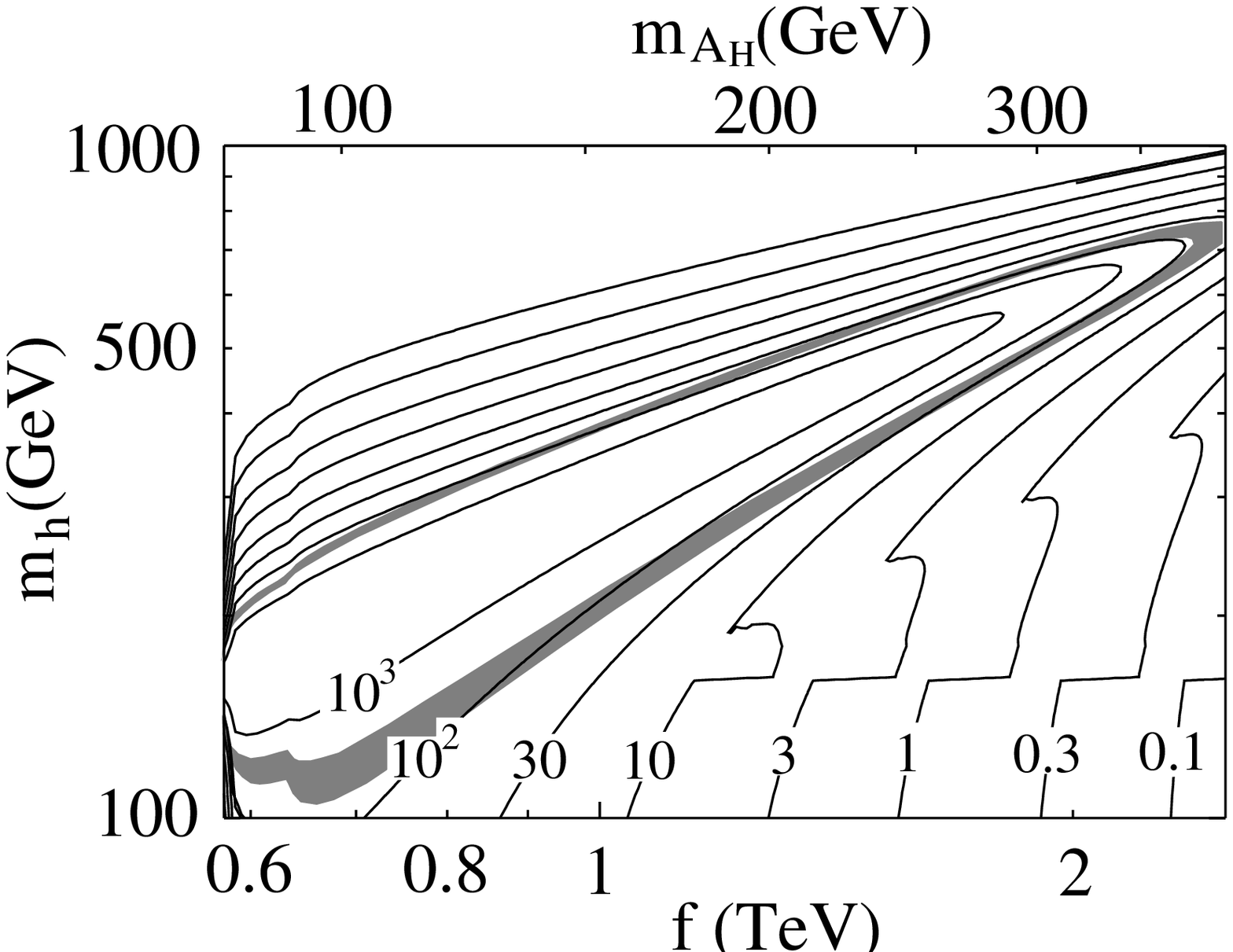}
  \caption{\small The contour plot of the $\chi^2$ in $(f,m_h)$-plane. The
   left figure is the $\chi^2$ in the PAMELA experiment with $BF = 5$, while
   the right one is the $\chi^2$ in the AMS-02 with $BF = 2$. For comparison,
   the constraint from the WMAP observation is also shown as a shaded region.
   The values of $\chi^2$, 30.8, 33.9 and 40.3, correspond to the
   statistical significance for the detection of the signal at the 90\%, 95\%
   and 99\% confidence levels.}
  \label{cont_chi2}
 \end{center}
\end{figure}

In Fig.\ref{cont_chi2}, the contour plot of the $\chi^2$ is depicted in
$(f,m_h)$-plane. The left figure is the $\chi^2$ in the PAMELA experiment
with $BF = 5$, while the right one is the $\chi^2$ in the AMS-02 with
$BF = 2$.\footnote{The value of $\chi^2$ is proportional to $BF^2$, thus
the extension of the result in other values of the boost factor is
straightforward.} The statistical significance for the detection of the
signal at the 90\%, 95\%
and 99\% confidence levels correspond to the values of $\chi^2$, 30.8, 33.9
and 40.3. In these figures, the constraint from the WMAP observation is also
shown as a shaded region. Inside the region, predicted thermal relics of the
dark matter are smaller, while the relics outside the region are larger than
that observed in the WMAP observation. If we like to consider parameter
region inconsistent with the WMAP observation, deviation from the thermal
relic scenario is necessary such as a non-thermal production of dark matter
or entropy production at late time.

The strange behavior around $f > 1$ TeV and $m_h \sim$ 200-400 GeV in the
figures is due to the annihilation mode into two Higgs bosons. When this
annihilation channel is opened ($m_{A_H} > m_h$) and the Higgs boson mass is
larger than twice the W boson mass ($m_H > 2m_W$), high energy positrons
are produced through the process $A_HA_H\rightarrow hh \rightarrow WWWW$.
These positrons have a hard spectrum, and enhance the possibility to detect
the dark matter signal. On the other hand, when the Higgs boson mass is less
than that of two W bosons ($m_H < 2m_W$), the effect becomes negligible.
Since positrons are produced through cascade decays of b-quarks, the
resultant spectrum in the process becomes very soft.

From the left figure, we see that the dark matter signal is clearly
distinguished from the background in the PAMELA experiment when the breaking
scale $f$ is less than 1 TeV for $BF = 5$. On the other hand, from the right
figure, almost all interesting area including the WMAP region is covered in
AMS-02 with $BF = 2$. Furthermore, the value of $\chi^2$ in the AMS-02 with
$BF = 1$ is quite similar to the plot in the PAMELA with $BF = 5$. Thus it is
possible to detect the signal in AMS-02 even if there is no enhancement
from the boost factor.

\begin{figure}[t]
 \begin{center}
  \includegraphics[width=7.3cm]{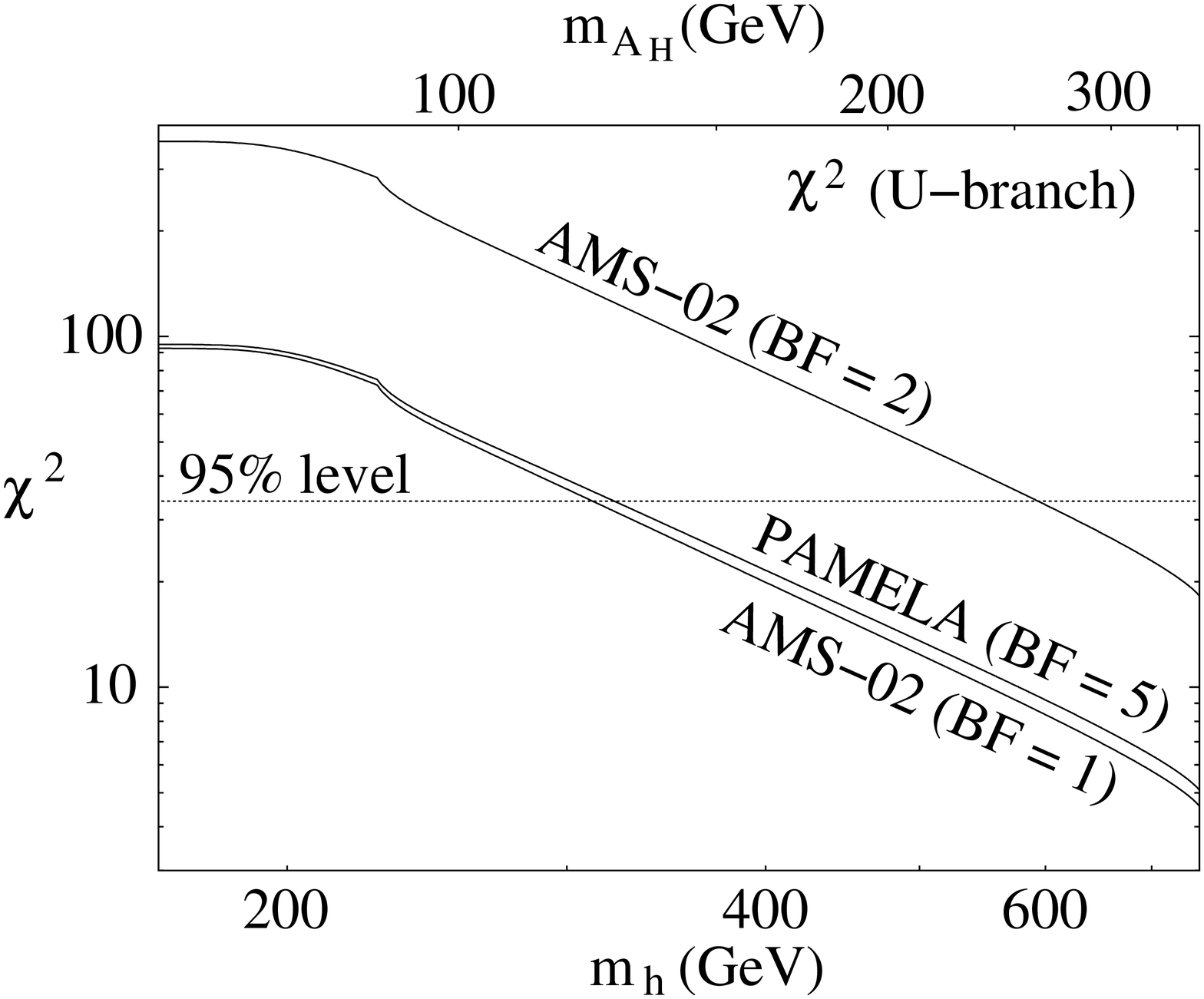}
  ~
  \includegraphics[width=7.3cm]{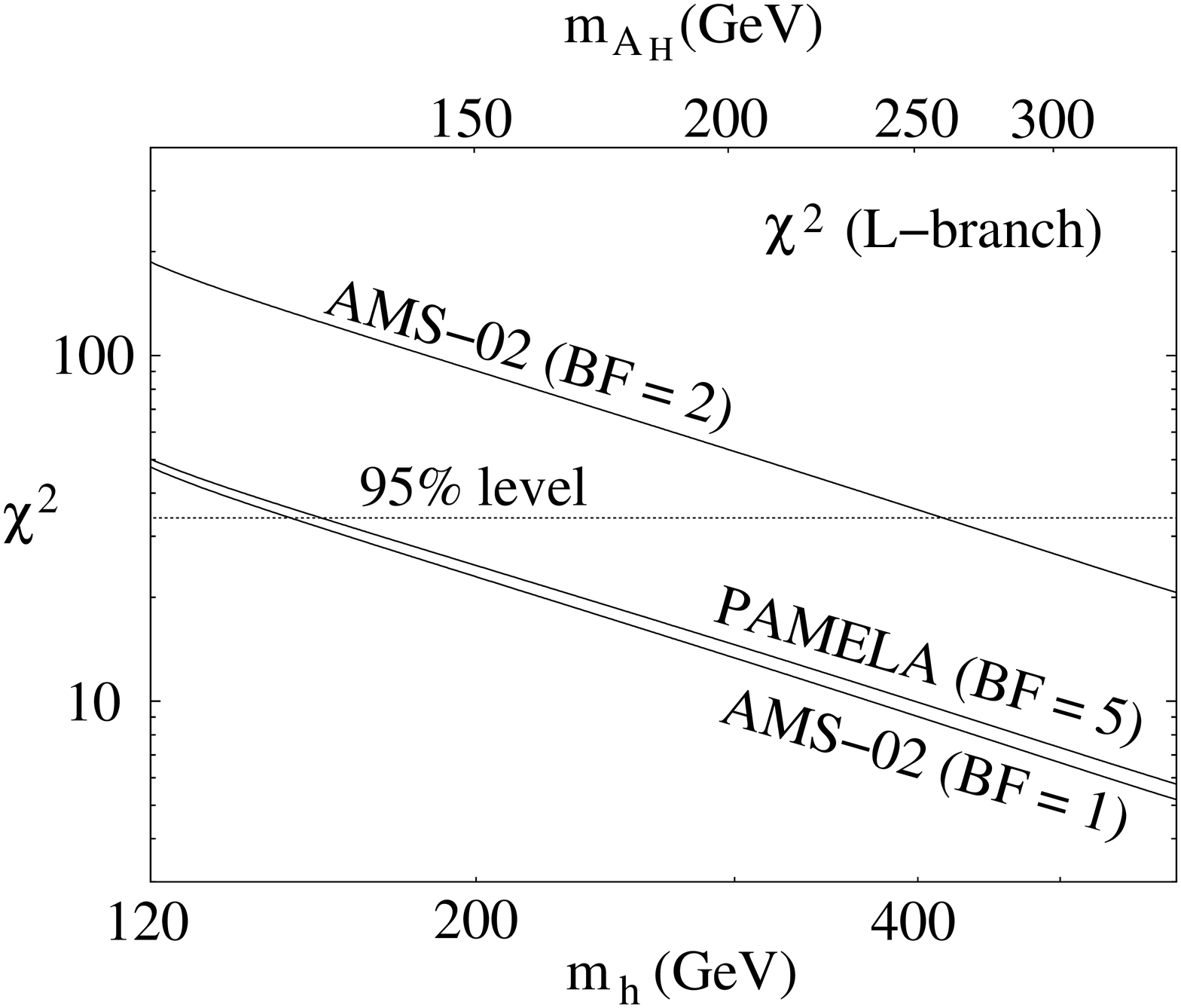}
  \caption{\small $\chi^2$ plot along the center value ($\Omega h^2 =
   0.112$) of U- (left figure) and L-branch (right figure) as a function of
   the Higgs boson mass $m_h$ and the dark matter mass $m_{A_H}$
   (see the top axis). In both figures, the $\chi^2$ of PAMELA with $BF = 5$,
   AMS-02 with $BF = 2$, and $BF = 1$ are depicted.}
  \label{chi2_wmap}
 \end{center}
\end{figure}

The $\chi^2$ plot along the center value ($\Omega h^2 = 0.112$) of
U- and L-branch as a function of the Higgs boson mass $m_h$ and dark matter
mass $m_{A_H}$ are presented in Fig.\ref{chi2_wmap}. The result of the
U-branch case is shown in the left figure, while that of the L-branch is in
the right figure. In both figures, the $\chi^2$ of the PAMELA with $BF = 5$,
AMS-02 with $BF = 2$ and $BF = 1$ are depicted. For the reference, the line
$\chi^2 = 33.9$ (corresponds to the 95\% confidence level) is shown.
The decreasing behavior of $\chi^2$ along with increasing $m_h$ is due to the
fact that a number density of the dark matter is decreasing as $m_{A_H}$ is
increasing. If the boost factor is around 5, the PAMELA experiment has a
potential to detect the dark matter signal when $m_h <$ 300 GeV ($m_{A_H}
< 120$ GeV) in the U-branch case or $m_h <$ 150 GeV ($m_{A_H} < 120$ GeV)
in the L-branch. Furthermore, these regions can be covered in AMS-02, even if
$BF = 1$.

\begin{figure}[t]
 \begin{center}
  \includegraphics[width=8cm]{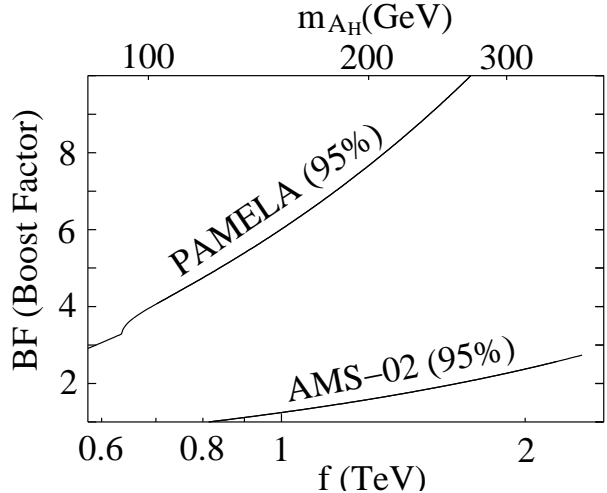}
  \caption{95\% confidence level for the statistical significance of the
   dark matter detection within the WMAP constraint in PAMELA and AMS-02
   experiments. The region above the line can be distinguished from the
   background spectrum. The plot is depicted as a function of the breaking
   scale $f$ (or the dark matter mass $m_{A_H}$ on the top axis) and the
   boost factor $BF$. The lines for U- and L-branches coincide.}
  \label{fig:chi2_cnt}
 \end{center}
\end{figure}

Finally, we show the 95\% confidence level contour within the WMAP constraint
in ($f$ or $m_{A_H}$,$BF$)-plane. The region above the line can be
distinguished from the background in each experiment. Although both results
in U- and L-branch cases are depicted in the figure, the contour lines are
almost degenerate in this parameter space. In both branches, the
annihilation modes are completely dominated by $WW$ and $ZZ$ bosons, and the
positron production cross sections are the same once the WMAP constraint is
applied. From the figure, we see that the signal may be detected in the PAMELA
experiment when $f < 830$ GeV ($m_{A_H} < 120$ GeV) and $BF > 5$. On the
other hand, the AMS-02 experiment will cover a wide range of the parameter
space including the region with $BF = 1$.

Here, we comment on the positron excess recently reported by HEAT
collaboration \cite{HEAT}. In the measurement, the excess of high energy
positrons (1 GeV $< E <$ 30 GeV) has been observed. If the excess is due to
the dark matter annihilation, its annihilation cross section should be large
($\sigma {\rm v} \sim 10^{-24}$cm$^3$sec$^{-1}$) unless the boost factor is
large ($BF\sim$ 50-100). Such large annihilation cross section is difficult
to satisfy the WMAP constraint\footnote{The cross section ($\sigma {\rm v}
\sim 2 \times 10^{-26}$cm$^3$sec$^{-1}$) is required in order to explain WMAP
result in the thermal relic scenario}. It is therefore unlikely that the
excess observed at the HEAT experiment is explained by the signal of the
dark matter in this model\footnote{There are large uncertainties on the
positron flux due to a solar modulation at GeV energy, which may be
responsible to the HEAT excess \cite{solarmod1}.}.

\vspace{1.0cm}
\lromn 5 \hspace{0.2cm} {\bf Summary and Discussions}
\vspace{0.5cm}

\begin{figure}[t]
 \begin{center}
  \includegraphics[width=8cm]{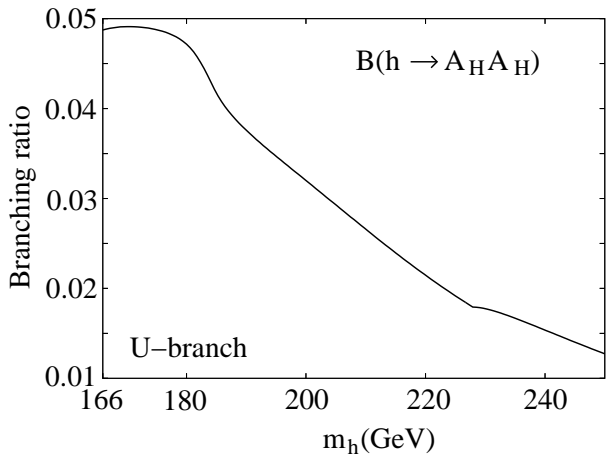}
  \caption{Branching ratio of the Higgs decay process into two dark matters
   along the center value of the U-branch as a function of the Higgs
   boson mass.}
  \label{fig:inv_width}
 \end{center}
\end{figure}

We have studied the possibility to detect the dark matter in the littlest
Higgs model with T-parity in future cosmic positron experiments. High energy
positrons are produced from the dark matter annihilation through weak gauge
boson decays. The resultant positron spectrum becomes hard, and the indirect
detection of the dark matter is promising. We have performed the
$\chi^2$-analysis to evaluate a confidence level to detect the dark matter
signal in upcoming experiments. We have found that the signal will be
detected in the PAMELA experiment, when the dark matter mass is less than
120 GeV and the boost factor is around 5 within the WMAP constraint. The
region $m_{A_H} <$ 120 GeV ($f <$ 830 GeV) corresponds to $m_h <$ 300 GeV in
the U-branch, and $m_h <$ 150 GeV in the L-branch. In the AMS-02
measurement, these regions can be covered even if there is no enhancement
from the boost factor.

The positron spectrum in this model has a different feature compared to a 
bino-like neutralino dark matter in supersymmetric models. In the
supersymmetric case, positrons are mostly produced from bottom quark decays, 
so that its spectrum becomes softer than the dark matter in the present
model.
Furthermore, the positron production cross section is much smaller, because
the bino-like dark matter annihilates in the p-wave. Therefore, the indirect
detection of the dark matter in the littlest Higgs model with T-parity is
easier than the supersymmetric case.

Finally, we discuss possible effects of the dark matter on the Higgs
phenomenology. In the U-branch, the decay of a Higgs boson into a pair of the
heavy photons contributes to the invisible width. The branching fraction of
the process is shown in Fig.\ref{fig:inv_width} as a function of the Higgs 
boson mass. We see that the ratio is at a few percent level. Although this
value seems to be beyond the currently estimated sensitivity to the total
width measurement of the Higgs boson in LHC \cite{Drollinger:2001bc} and
ILC \cite{Meyer:2004ha}, the measurement might be possible at a future muon
collider \cite{Alsharoa:2002wu}.

We have concentrated on the positron signal of the dark matter in this paper.
It is also interesting to consider other ways to search for the dark matter
signals such as direct detection, and indirect detections using neutrinos,
gamma-rays and anti-protons.

\vspace{1.0cm}
\hspace{0.2cm} {\bf Acknowledgments}
\vspace{0.5cm}

This work is supported in part by the Grant-in-Aid for Science Research, 
Ministry of Education, Science and Culture, Japan (No.16081211 for S.M and
Y.O, No.15740164 for N.O., and Nos.13135225 and 17540286 for Y.O.).

\vspace{1.0cm}
{\bf Appendix}\hspace{0.2cm} {\bf Constraints from Electroweak Precision
Measurements}
\vspace{0.5cm}

In Appendix, we consider constraints on the Littlest Higgs model with T-parity
from electroweak precision measurements. We follow the procedure in
Ref.\cite{Hubisz:2005tx} using S, T and U parameters \cite{Peskin:1991sw}. In
that paper, it has been shown that main contributions to these parameters come
from the top-sector in the model and custodial-symmetry violating effects from
heavy gauge boson loops.

Contribution from the top-sector depends not only on the parameter $f$ but also
on a new parameter, $R \equiv \lambda_1/\lambda_2$. Using $\lambda_1$ and
$\lambda_2$, masses of the top quark and its T-even partner are given as
$m_t = \lambda_1\lambda_2v/(\lambda_1^2 + \lambda_2^2)^{1/2}$ and $m_{T_+} =
(\lambda_1^2 + \lambda_2^2)^{1/2}f$. Instead of $\lambda_1$, $\lambda_2$, and
$f$, we take $m_t$, $R$, and $f$ as free parameters. We have calculated
contributions from the top sector to the S, T and U parameters, and confirm
the results in Ref.\cite{Hubisz:2005tx}, which are given by
\begin{eqnarray}
 S_T
 &=&
 \frac{1}{3\pi}
 \left(\frac{\lambda_1}{\lambda_2}\right)^2
 \frac{m_t^2}{m_{T_+}^2}
 \left[
  -
  \frac{5}{2}
  +
  \log\frac{m_{T_+}^2}{m_t^2}
 \right]~,
 \nonumber \\
 T_T
 &=&
 \frac{3}{8\pi}
 \frac{1}{s_W^2c_W^2}
 \left(\frac{\lambda_1}{\lambda_2}\right)^2
 \frac{m_t^4}{m_{T_+}^2 m_Z^2}
 \left[
  \log \frac{m_{T_+}^2}{m_t^2}
  -
  1
  +
  \frac{1}{2}\left(\frac{\lambda_1}{\lambda_2}\right)^2
 \right]~,
 \nonumber \\
 U_T
 &=&
 \frac{5}{6\pi}
 \left(\frac{\lambda_1}{\lambda_2}\right)^2
 \frac{m_t^2}{m_{T_+}^2}~.
 \label{Tcontribution}
\end{eqnarray}

We have also calculated contributions from heavy gauge boson loops in the
general $R_\xi$ gauge. We have obtained $\xi$ independent results. Final 
expression, however, is not consistent with the result in
Ref.\cite{Hubisz:2005tx}, in which the Landau gauge has been used. Detail of
our calculation is presented in the followings.

In order to define the $R_\xi$ gauge, we first consider the mixing term between
gauge bosons and derivatives of Nambu-Goldstone (NG) bosons up to appropriate
order of $v/f$. The non-linear sigma field $\Sigma$ is expanded by the NG
fields as
\begin{eqnarray}
 \Sigma
 &\equiv&
 \exp\left[2i(\langle\Pi\rangle + \delta\Pi)/f\right]\Sigma_0
 \nonumber \\
 &=&
 ({\rm constant})
 +
 ({\rm terms} \propto \delta\Pi)
 +
 ({\rm terms} \propto \delta\Pi^2)
 +
 \cdots~,
\end{eqnarray}
where
\begin{eqnarray}
 \langle\Pi\rangle
 &=&
 \left(
  \begin{array}{ccccc}
   0 &   0 &   0 & 0 &   0 \\
   0 &   0 & v/2 & 0 &   0 \\
   0 & v/2 &   0 & 0 & v/2 \\
   0 &   0 &   0 & 0 &   0 \\
   0 &   0 & v/2 & 0 &   0 \\
  \end{array}
 \right)~,
 \\
 \delta\Pi
 &=&
 \left(
  \begin{array}{ccccc}
    -\frac{\omega^0}{2} - \frac{\eta}{\sqrt{20}}
  & -\frac{\omega^\dagger}{\sqrt{2}}
  & -\frac{i\pi^\dagger}{\sqrt{2}}
  & -i\phi^{++}
  & -\frac{i\phi^\dagger}{\sqrt{2}} \\
    -\frac{\omega}{\sqrt{2}}
  &  \frac{\omega^0}{2} - \frac{\eta}{\sqrt{20}}
  &  \frac{h + i\pi^0}{2}
  & -\frac{i\phi^\dagger}{\sqrt{2}}
  &  \frac{-i\phi^0 +\phi_P^0}{\sqrt{2}} \\
     \frac{i\pi}{\sqrt{2}}
  &  \frac{h - i\pi^0}{2}
  &  \frac{2\eta}{\sqrt{5}}
  & -\frac{i\pi^\dagger}{\sqrt{2}}
  &  \frac{h+i \pi^0}{2} \\
    i\phi^{--}
  &  \frac{i\phi}{\sqrt{2}}
  &  \frac{i\pi}{\sqrt{2}}
  & -\frac{\omega^0}{2} - \frac{\eta}{\sqrt{20}}
  & -\frac{\omega}{\sqrt{2}} \\
     \frac{i\phi}{\sqrt{2}}
  &  \frac{i \phi^0 +\phi_P^0}{\sqrt{2}}
  &  \frac{h - i\pi^0}{2}
  & -\frac{\omega^\dagger}{\sqrt{2}}
  &  \frac{\omega^0}{2} - \frac{\eta}{\sqrt{20}}
  \end{array}
 \right)~.
 \nonumber
\end{eqnarray}
We used the notation adopted in Ref.\cite{Hubisz:2005tx} for the NG fields
in the $\delta\Pi$ matrix. It is well known that the constant term is obtained
in an exact manner as
\begin{eqnarray}
 ({\rm constant})
 &=&
 \left(
  \begin{array}{ccccc}
   0 &           0 &           0 & 1 &           0 \\
   0 &  -(1 - c)/2 & is/\sqrt{2} & 0 &   (1 + c)/2 \\
   0 & is/\sqrt{2} &           c & 0 & is/\sqrt{2} \\
   1 &           0 &           0 & 0 &           0 \\
   0 &   (1 + c)/2 & is/\sqrt{2} & 0 &  -(1 - c)/2 \\
  \end{array}
 \right)~,
 \label{constant}
\end{eqnarray}
where $s = \sin(\sqrt{2}v/f)$, $c = \cos(\sqrt{2}v/f)$. Thanks to the Feynman
formula \cite{Feynman formula}, we can also derive terms proportional to
$\delta \Pi$ exactly as follows,
\begin{eqnarray}
 ({\rm terms} \propto \delta\Pi)
 &=&
 \int^1_0 d\alpha~
 e^{2i(1 - \alpha)\langle\Pi\rangle/f}
 \left(\frac{2\delta\Pi i}{f}\right)
 e^{2i\alpha\langle\Pi\rangle/f}
 \Sigma_0~.
 \label{first}
\end{eqnarray}
The integration by the parameter $\alpha$ can be exactly performed. Due to the
complex form of the result, we omit to write it down. Using these exact
expressions in Eqs.(\ref{constant}) and (\ref{first}), the mixing term induced
from the kinetic term in Eq.(\ref{Kinetic L}) is written as
\begin{eqnarray}
 \frac{g f^2(1 - c)}{2v}W^a\partial\pi^a
 +
 \left(
 \frac{g f^2}{2}W_H\cdot
 \left[
   \left(\frac{1}{f} + \frac{s}{\sqrt{2}v}\right)\partial\omega^\dagger
  -
  i\left(\frac{1}{f} - \frac{s}{\sqrt{2}v}\right)\partial\phi^\dagger
 \right]
 +
 h.c.
 \right)
 \nonumber \\
 +
 \frac{g f^2}{8}Z_H\cdot
 \left[
          \left(\frac{7}{f} + \frac{s c}{\sqrt{2}v}\right)\partial\omega^0
  -
  \sqrt{2}\left(\frac{1}{f} - \frac{s c}{\sqrt{2}v}\right)\partial\phi_P^0
  +
  \sqrt{5}\left(\frac{1}{f} - \frac{s c}{\sqrt{2}v}\right)\partial\eta
 \right]~,
 \label{mixing}
\end{eqnarray}
where $W_H = (W_H^1 + iW_H^2)/\sqrt{2}$, and $Z_H = W_H^3$. We neglect the
effect of U(1) gauge interactions for simplicity ($g' = 0$).

Due to the electroweak symmetry breaking, the NG mode absorbed in the
longitudinal component of the heavy gauge boson $W_H$ ($Z_H$) is given by the
combination of $\omega$ and $\phi$ ($\omega^0$, $\phi_P^0$, and $\eta$). Thus,
``would-be NG'' modes are written as
\begin{eqnarray}
 \tilde{\pi}^a
 &=&
 N_{\tilde{\pi}}\pi^a~,
 \label{WBNG}
 \\
 \tilde{\omega}
 &=&
 N_{\tilde{\omega}}
 \left[
   \left(1 + \frac{f s}{\sqrt{2}v}\right)\omega
  +
  i\left(1 - \frac{f s}{\sqrt{2}v}\right)\phi
 \right]~,
 \nonumber \\
 \tilde{\omega}^0
 &=&
 N_{\tilde{\omega}^0}
 \left[
          \left(7 + \frac{f s c}{\sqrt{2}v}\right)\omega^0
  -
  \sqrt{2}\left(1 - \frac{f s c}{\sqrt{2}v}\right)\phi_P^0
  +
  \sqrt{5}\left(1 - \frac{f s c}{\sqrt{2}v}\right)\eta
 \right]~.
 \nonumber
\end{eqnarray}
Note that these modes become $\pi^a$, $\omega$, and $\omega^0$ at the leading
order of $v/f$. Normalization constants, $N_x$ ($x = \tilde{\pi}$,
$\tilde{\omega}$, and $\tilde{\omega}^0$), are determined by considering
kinetic terms of NG bosons. The kinetic terms are also obtained exactly as
\begin{eqnarray}
 &&
 \frac{f^2(1 - c)}{2v^2}\left(\partial\pi^a\right)^2
 +
 \frac{1}{2}\left(\partial h\right)^2
 +
 \partial \phi^{++} \partial \phi^{--}
 +
 \frac{f^2(1 - c)}{2v^2}\left(\partial \phi^0\right)^2
 \\
 &+&
 \left(\frac{(1 - c)f^2}{2v^2} + \frac{1}{2}\right)
 \left(
  \partial\phi^\dagger\partial\phi
  +
  \partial\omega^\dagger\partial\omega
 \right)
 +
 i\left(\frac{(1 - c)f^2}{2v^2} - \frac{1}{2}\right)
 \left(
  \partial\phi^\dagger\partial\omega
  -
  \partial\omega^\dagger\partial\phi
 \right)
 \nonumber \\
 &+&
 \frac{1}{16}\left(7 + \frac{s^2f^2}{2v^2}\right)
 \left(\partial\omega^0\right)^2
 +
 \frac{1}{16}\left(3 + \frac{5s^2f^2}{2v^2}\right)
 \left(\partial\eta\right)^2
 +
 \frac{1}{8}\left(3 + \frac{s^2f^2}{2v^2}\right)
 \left(\partial\phi_P^0\right)^2
 \nonumber \\
 &+&
 \frac{\sqrt{5}}{8}\left(1 - \frac{s^2f^2}{2v^2}\right)
 \partial\omega^0\partial\eta
 -
 \frac{\sqrt{2}}{8}\left(1 - \frac{s^2f^2}{2v^2}\right)
 \partial\omega^0\partial\phi_P^0
 +
 \frac{\sqrt{10}}{8}\left(1 - \frac{s^2f^2}{2v^2}\right)
 \partial\eta\partial\phi_P^0~.
 \nonumber
\end{eqnarray}
As seen from the equation, we find that kinetic terms of $\pi$, $\omega$,
$\phi$, $\omega^0$, $\phi_P^0$, $\eta$, and $\phi^0$ are not canonically
normalized due to the electroweak symmetry breaking. Therefore, we have to
redefine these NG fields by normalized ones. After the redefinition, we found
the normalization constants, $N_x$, in Eq.(\ref{WBNG}) are given as
\begin{eqnarray}
 N_{\tilde{\pi}}
 =
 \frac{f\sqrt{1 - c}}{v}~,
 \qquad
 N_{\tilde{\omega}}
 = 
 \frac{1}{\sqrt{3 + c}}~,
 \qquad
 N_{\tilde{\omega}^0}
 =
 \frac{1}{2\sqrt{14 + 2 c^2}}~.
\end{eqnarray}

Once we obtain canonically normalized ``would-be NG'' modes, we can determine
gauge fixing functions to cancel the mixing term in Eq.(\ref{mixing}). Those
functions are
\begin{eqnarray}
 &&
 G^a
 =
 \frac{1}{\sqrt{\xi}}
 \left[
  \partial W^a
  -
  \xi
  \frac{g f\sqrt{1 - c}}{2}\tilde{\pi}^a
 \right]~,
 \label{gauge_fixing}
 \\
 &&
 G^{1,2}_H
 =
 \frac{1}{\sqrt{\xi}}
 \left[
  \partial W^a_H
  -
  \xi
  \frac{g f\sqrt{3 + c}}{2}\tilde{\omega}^a
 \right]~,
 \qquad
 G^3_H
 =
 \frac{1}{\sqrt{\xi}}
 \left[
  \partial Z_H
  -
  \xi
  \frac{g f\sqrt{7 + c^2}}{2\sqrt{2}}\tilde{\omega}^0
 \right]~.
 \nonumber
\end{eqnarray}
Using these functions, mass terms of the ``would-be NG'' modes turn out to be
\begin{eqnarray}
 \left.
  \frac{\left(G^a\right)^2 + \left(G_H^a\right)^2}{2}
 \right|_{\rm NG~mass}
 =
 \frac{\xi}{2}m_W^2
 \left(\tilde{\pi}^a\right)^2
  +
  \xi m_{W_H}^2\tilde{\omega}^\dagger \tilde{\omega}
  +
  \frac{\xi}{2}\left(m_{W_H}^2 + \Delta M^2\right)
  \left(\tilde{\omega}^0\right)^2~,
\end{eqnarray}
where $m_W^2 = g^2f^2(1 - c)/4$, $m_{W_H}^2 = g^2f^2(3 + c)/4$, and $\Delta M^2 =
g^2f^2(1 - c)^2/8$. Hence, with $\xi = 1$, ``would-be NG'' masses coincide
with those of corresponding gauge bosons at any order of $v/f$.

We are now in position to calculate the contribution to the T parameter from
heavy gauge boson loops. With expressions of ``would-be NG'' modes in
Eq.(\ref{WBNG}), interactions relevant to the calculation are given as follows,
\begin{eqnarray}
 &&
 {\cal L}
 =
 {\cal L}_{\rm V_L V_H V_H}
 +
 {\cal L}_{\rm V_L V_L V_H V_H}
 +
 {\cal L}_{\rm Ghost}
 +
 {\cal L}_{\rm V_L NG NG}
 +
 {\cal L}_{\rm V_L V_L NG NG}
 +
 {\cal L}_{\rm V_L V_H NG}~,
 \nonumber
\end{eqnarray}
\begin{eqnarray}
 {\cal L}_{\rm V_L V_H V_H}
 &=&
 i g
 \left[~~~~
  Z_\mu W_{H\nu}^\dagger
  \left(
   \partial^\mu W_H^\nu - \partial^\nu W_H^\mu
  \right)
  -
  Z_\mu W_{H\nu}
  \left(
   \partial^\mu W_H^{\dagger\nu} - \partial^\nu W_H^{\dagger\mu}
  \right)
 \right.
 \nonumber \\
 &&\qquad
 +
 \partial_\mu Z_\nu
 \left(
  W_H^{\dagger\mu}W_H^\nu - W_H^\mu W _H^{\dagger\nu}
 \right)
 +
 W^\dagger_\mu W_{H\nu}
 \left(
  \partial^\mu Z_H^\nu - \partial^\nu Z_H^\mu
 \right)
 \nonumber \\
 &&\qquad
 -
 W^\dagger_\mu Z_{H\nu}
 \left(
  \partial^\mu W_H^\nu - \partial^\nu W_H^\mu
 \right)
 +
 \partial_\mu W^\dagger_\nu
 \left(
  Z_H^\nu W_H^\mu - Z_H^\mu W_H^\nu
 \right)
 \nonumber \\
 &&\qquad
 -
 W_\mu W^\dagger_{H\nu}
 \left(
  \partial^\mu Z_H^\nu - \partial^\nu Z_H^\mu
 \right)
 +
 W_\mu Z_{H\nu}
 \left(
  \partial^\mu W_H^{\dagger\nu} - \partial^\nu W_H^{\dagger\mu}
 \right)
 \nonumber \\
 &&\qquad
 \left.
  -
  \partial_\mu W_\nu
  \left(
   Z_H^\nu W_H^{\dagger\mu} - Z_H^\mu W_H^{\dagger\nu}
  \right)~~~
 \right]~,
\end{eqnarray}
\begin{eqnarray}
 {\cal L}_{\rm V_L V_L V_H V_H}
 &=&
 -\frac{g^2}{2}
 \left[
  f^{a b e}f^{c d e}g^{\mu\mu'}g^{\nu\nu'}
  +
  f^{a d e}f^{c b e}g^{\mu\mu'}g^{\nu\nu'}
  +
  f^{a c e}f^{b d e}g^{\mu\nu}g^{\mu'\nu'}
 \right]
 \nonumber \\
 &&\qquad
 \times
 W_\mu^a W_\nu^b W_{H\mu'}^c W_{H\nu'}^d~,
\end{eqnarray}
\begin{eqnarray}
 {\cal L}_{\rm Ghost}
 =
 g f^{a b c}\bar{\eta}^a_H
 \partial^\mu
 \left(
  W^b_\mu \eta^c_H
 \right)~,
\end{eqnarray}
\begin{eqnarray}
 {\cal L}_{\rm V_L NG NG}
 &=&
 \frac{g}{2}
 \left[
  \frac{3c + 5}{c + 3}
  Z\cdot
  \left(
   \tilde{\omega}i\partial\tilde{\omega}^\dagger
   -
   \tilde{\omega}^\dagger i\partial\tilde{\omega}
  \right)
 \right.
 \nonumber \\
 &&\qquad
 \left.
  +
  \frac{c^2 + 4c + 11}{\sqrt{2(c + 3)(c^2 + 7)}}
  \left\{
   W\cdot
   \left(
    \tilde{\omega}^\dagger i\partial\tilde{\omega}^0
    -
    \tilde{\omega}^0 i\partial\tilde{\omega}^\dagger
   \right)
   +
   h.c.
   \right\}
 \right]
 \\
 &=&
 g
 \left(
  1 - \frac{v^2}{8f^2} - \frac{v^4}{96f^4}
 \right)
 \left[
  Z\cdot
  \left(
   \tilde{\omega}i\partial\tilde{\omega}^\dagger
   -
   \tilde{\omega}^\dagger i\partial\tilde{\omega}
  \right)
 \right.
 \nonumber \\
 &&\qquad\qquad\qquad
 \left.
  +
  \left\{
   W\cdot
   \left(
    \tilde{\omega}^\dagger i\partial\tilde{\omega}^0
    -
    \tilde{\omega}^0 i\partial\tilde{\omega}^\dagger
   \right)
   +
   h.c.
  \right\}
 \right]
 +
 {\cal O}\left(\frac{v^6}{f^6}\right)~,
 \nonumber
\end{eqnarray}
\begin{eqnarray}
 {\cal L}_{\rm V_L V_L NG NG}
 &=&
 g^2
 \left[
  C_{\tilde{\omega}^0\tilde{\omega}^0ZZ}
  \left(\tilde{\omega}^0\right)^2Z_\mu Z^\mu
  +
  C_{\tilde{\omega}^\dagger\tilde{\omega}ZZ}
  \tilde{\omega}^\dagger\tilde{\omega}Z_\mu Z^\mu
 \right.
 \nonumber \\
 &&\qquad
 \left.
  +
  C_{\tilde{\omega}^\dagger\tilde{\omega}W^\dagger W}
  \tilde{\omega}^\dagger\tilde{\omega}W^\dagger_\mu W^\mu
  +
  C_{\tilde{\omega}^0\tilde{\omega}^0W^\dagger W}
  \left(\tilde{\omega}^0\right)^2W^\dagger_\mu W^\mu
 \right]~,
 \label{qurtic}
\end{eqnarray}
\begin{eqnarray}
 {\cal L}_{\rm V_L V_H NG}
 &=&
 i g^2f
 \left[
  \frac{\sqrt{2}(1 + c)}{\sqrt{c^2 + 7}}
  \tilde{\omega}^0W^\dagger_\mu W_H^\mu
  +
  \frac{c^2 + 2c + 5}{4\sqrt{c + 3}}
  \tilde{\omega}^\dagger W_\mu Z_H^\mu
 \right.
 \nonumber \\
 &&\qquad\qquad\qquad\qquad\qquad\qquad\qquad
 +
 \left.
  \frac{1 + c}{\sqrt{c + 3}}
  \tilde{\omega} W_{H\mu}^\dagger Z^\mu
 \right]
 +
 h.c.
 \nonumber \\
 &=&
 i g^2f
 \left[
  \left(
   1
   -
   \frac{3v^2}{8f^2}
   -
   \frac{5v^4}{128f^4}
  \right)
  \tilde{\omega}^0W^\dagger_\mu W_H^\mu
  +
  \left(
   1
   -
   \frac{3v^2}{8f^2}
   +
   \frac{19v^4}{128f^4}
  \right)
  \tilde{\omega}^\dagger W_\mu Z_H^\mu
 \right.
 \nonumber \\
 &&\qquad\qquad
 +
 \left.
  \left(
   1
   -
   \frac{3v^2}{8f^2}
   +
   \frac{3v^4}{128f^4}
  \right)
  \tilde{\omega} W_{H\mu}^\dagger Z^\mu
 \right]
 +
 h.c.
 +
 {\cal O}\left(\frac{v^6}{f^6}\right)~.
 \label{key}
\end{eqnarray}
In order to calculate the coefficients, C,  in Eq.(\ref{qurtic}), we need not
only the linear terms in Eq.(\ref{first}), but also those proportional to
$\delta\Pi^2$. Since it is difficult to obtain the $\delta\Pi^2$ terms in an
exact manner, we have calculated these terms with the expansion of $v/f$. Up to
the order of $(v/f)^6$, the coefficients turn out to be
\begin{eqnarray}
 C_{\tilde{\omega}^0\tilde{\omega}^0ZZ}
 &=&
 ~~
 -
 \frac{v^2}{96 f^2}
 +
 \frac{37 v^4}{5760 f^4}
 +
 {\cal O}\left(\frac{v^6}{f^6}\right)~,
 \nonumber \\
 C_{\tilde{\omega}^\dagger\tilde{\omega}ZZ}
 &=&
 1
 -
 \frac{13 v^2}{48 f^2}
 -
 \frac{47 v^4}{2880 f^4}
 +
 {\cal O}\left(\frac{v^6}{f^6}\right)~,
 \nonumber \\
 C_{\tilde{\omega}^\dagger\tilde{\omega}W^\dagger W}
 &=&
 1
 -
 \frac{7 v^2}{24 f^2}
 +
 \frac{163 v^4}{1440 f^4}
 +
 {\cal O}\left(\frac{v^6}{f^6}\right)~,
 \nonumber \\
 C_{\tilde{\omega}^0\tilde{\omega}^0W^\dagger W}
 &=&
 1
 -
 \frac{13 v^2}{48 f^2}
 -
 \frac{383 v^4}{2880 f^4}
 +
 {\cal O}\left(\frac{v^6}{f^6}\right)~.
\end{eqnarray}
Interactions in ${\cal L}_{\rm V_L V_H V_H}$ and
${\cal L}_{\rm V_L V_L V_H V_H}$ are gauge self-interactions coming from gauge
kinetic terms, while ghost interactions in ${\cal L}_{\rm Ghost}$ are derived
from the gauge fixing functions in Eq.(\ref{gauge_fixing}). Interactions in
${\cal L}_{\rm V_L NG NG}$, ${\cal L}_{\rm V_L V_L NG NG}$ and
${\cal L}_{\rm V_L V_H NG}$ have been obtained by expanding the kinetic term in
Eq.(\ref{Kinetic L}) in terms of NG fields.

\begin{figure}[t]
\begin{center}
\scalebox{0.68}{\includegraphics*{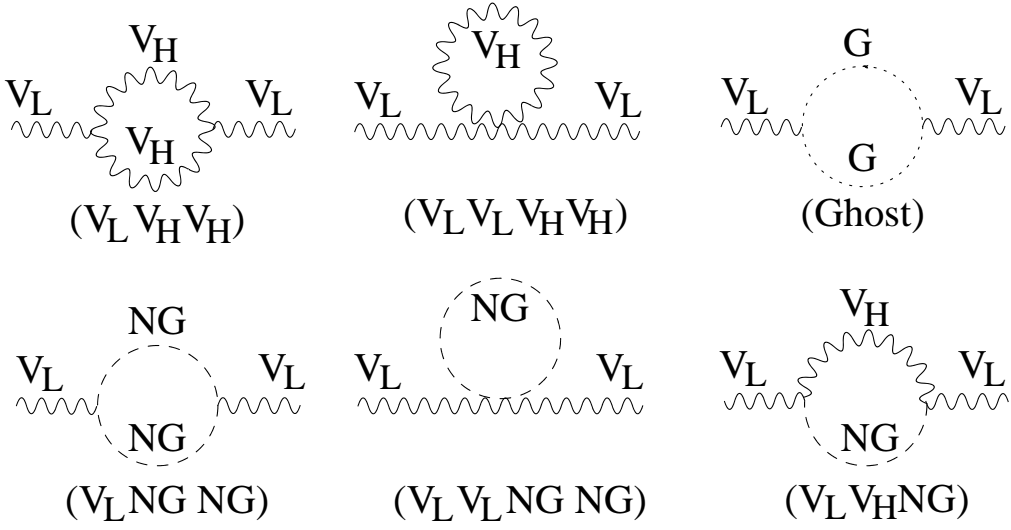}}
\caption{\small Heavy gauge boson loop diagrams contributing to the
         T parameter.}
\label{fig:GH_loops}
\end{center}
\end{figure}

There are six kinds of loop diagrams involving heavy gauge and NG bosons
contributing to the T parameter as shown in Fig.\ref{fig:GH_loops}. Those are
diagrams with three points gauge self-interactions (${\rm V_L V_H V_H}$), four
points gauge self-interactions (${\rm V_L V_L V_H V_H}$), ghost interactions
(Ghost), three points NG gauge interactions (${\rm V_L NG NG}$), four points NG
gauge interactions (${\rm V_L V_L NG NG}$), and interactions in
${\cal L}_{\rm V_L V_H NG}$ (${\rm V_L V_H NG}$). In all diagrams except the
last one, the contributions to the T parameter come from the mass difference
between $W_H$ and $Z_H$. In the last diagram, difference among coefficients
in Eq.(\ref{key}) is the source of the T parameter.

The leading logarithmic divergent contribution from each diagram is
\begin{eqnarray}
 T({\rm V_L V_H V_H})
 &=&
 \frac{4\pi}{s_W^2c_W^2m_Z^2}
 \left(\frac{\Delta M^2}{16\pi^2}\right)
 \left(3 + \frac{3}{4}\xi + \frac{3}{4}\xi^2\right)
 \ln\left(\frac{\Lambda^2}{m_{W_H}^2}\right)~,
 \nonumber \\
 T({\rm V_L V_L V_H V_H})
 &=&
 \frac{4\pi}{s_W^2c_W^2m_Z^2}
 \left(\frac{\Delta M^2}{16\pi^2}\right)
 \left(-\frac{9}{4} - \frac{3}{4}\xi^2\right)
 \ln\left(\frac{\Lambda^2}{m_{W_H}^2}\right)~,
 \nonumber \\
 T({\rm Ghost})
 &=&
 \frac{4\pi}{s_W^2c_W^2m_Z^2}
 \left(\frac{\Delta M^2}{16\pi^2}\right)
 \left(-\frac{1}{2}\xi\right)
 \ln\left(\frac{\Lambda^2}{m_{W_H}^2}\right)~,
 \nonumber \\
 T({\rm V_L NG NG})
 &=&
 \frac{4\pi}{s_W^2c_W^2m_Z^2}
 \left(\frac{\Delta M^2}{16\pi^2}\right)
 \left(+\xi\right)
 \ln\left(\frac{\Lambda^2}{m_{W_H}^2}\right)~,
 \nonumber \\
 T({\rm V_L V_L NG NG})
 &=&
 \frac{4\pi}{s_W^2c_W^2m_Z^2}
 \left(\frac{\Delta M^2}{16\pi^2}\right)
 \left(-\xi\right)
 \ln\left(\frac{\Lambda^2}{m_{W_H}^2}\right)~,
 \nonumber \\
 T({\rm V_L V_H NG})
 &=&
 \frac{4\pi}{s_W^2c_W^2m_Z^2}
 \left(\frac{\Delta M^2}{16\pi^2}\right)
 \left(- \frac{3}{4} - \frac{1}{4}\xi\right)
 \ln\left(\frac{\Lambda^2}{m_{W_H}^2}\right)~.
\end{eqnarray}
After summing all contributions, we have found that the logarithmic divergent
correction is completely canceled,
\begin{eqnarray}
 T_{\rm V_H}
 =
 0
 \times
 \frac{4\pi}{s_W^2c_W^2m_Z^2}
 \left(\frac{\Delta M^2}{16\pi^2}\right)
 \ln\left(\frac{\Lambda^2}{m_{W_H}^2}\right)
 +
 ({\rm finite~terms})~.
 \label{zero}
\end{eqnarray}
This result differs from Eq.(3.6) in Ref.\cite{Hubisz:2005tx}.
Therefore, the contribution to the T parameter at this order of $v/f$ comes
from a finite term.

\begin{figure}[t]
\begin{center}
\scalebox{0.65}{\includegraphics*{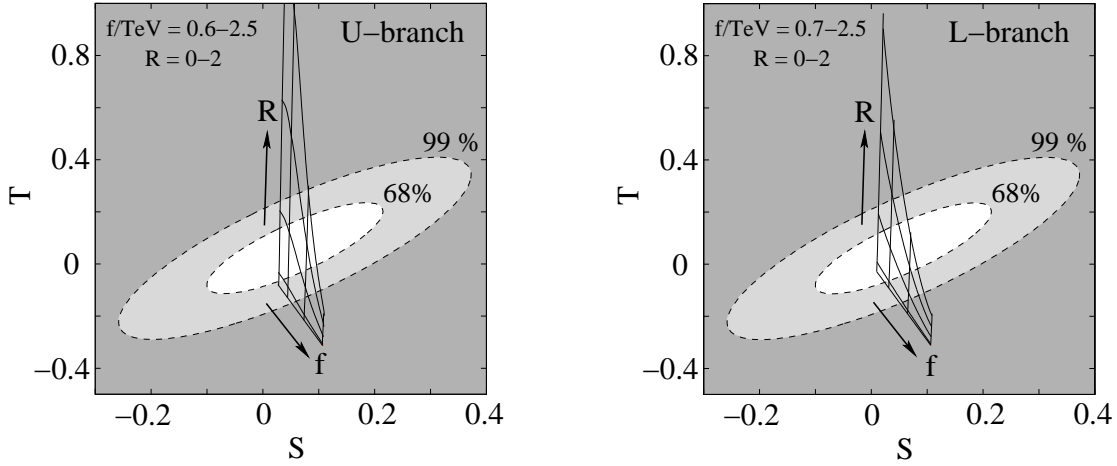}}
\caption{\small Contributions to S and T parameters in the Littlest Higgs
         model with T-parity (solid lines) for $0 < R < 2$ and $f$ along the
         center value of the U-branch (left figure) and L-branch
         (right figure). From bottom to top, lines correspond to $R =$ 0, 0.5,
         1, 1.5 and 2. From left to right, $f$ corresponds to the point I, II,
         III, and VII for U-branch, IV, V, VI, and VII for L-branch in
         Fig.\ref{fig:sample_points}. Constraints on S and T parameters from
         electroweak precision measurements are also shown at 68\% and 99\%
         confidence level. The plot assumes $U = 0$.}
\label{ST}
\end{center}
\end{figure}

In addition to the loop contribution, we expect that the following operator can
arise at the cutoff scale \cite{Hubisz:2005tx},
\begin{eqnarray}
 {\cal L}_c
 =
 \delta_c\frac{g^2}{16\pi^2}f^2
 {\rm Tr}
 \left[
  (Q_i^a D_\mu \Sigma)(Q_i^a D^\mu \Sigma)^*
 \right]~,
\end{eqnarray}
which gives the contribution to the T parameter at the same order of the finite
term,
\begin{eqnarray}
 T({\rm Cutoff})
 =
 \frac{4\pi}{s_W^2c_W^2m_Z^2}
 \left(\frac{\Delta M^2}{16\pi^2}\right)
 \left(- \delta_c\right)~.
 \label{TCutoff}
\end{eqnarray}
The value of $\delta_c$ is determined by the UV-completion of the Littlest
Higgs model with T-parity, and it is naturally expected to be ${\cal O}(1)$.
By comparing the contribution from the top sector with those from heavy gauge
boson loops and the cutoff scale operator, the latter contributions turns out
to be small as long as the finite term in Eq.(\ref{zero}) does not have a large
coefficient. In fact, $T$(Cutoff) in Eq.(\ref{TCutoff}) is 8\% of $T_T$ in 
Eq.(\ref{Tcontribution}), when we take $f = 1$ TeV, $R = 1$, and $\delta_c =
1$. Therefore, we neglect these contributions to derive electroweak precision
constraints on the model.

\begin{figure}[t]
\begin{center}
\scalebox{0.84}{\includegraphics*{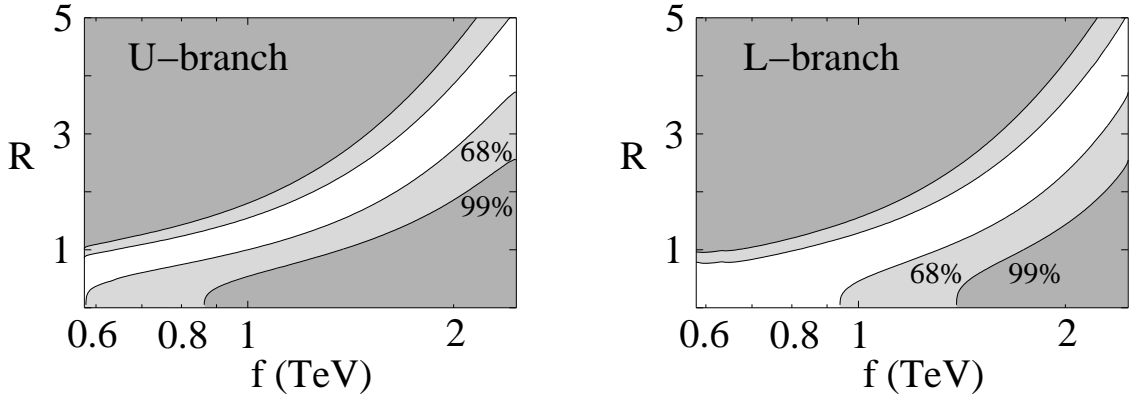}}
\caption{\small Constraints on the model parameters, $R$ and $f$, at 68\% and
         99\% confidence level. At each point in these figures, the Higgs mass
         is determined to satisfy the WMAP constraint on the U-branch (left
         figure) and L-branch (right figure).}
\label{STII}
\end{center}
\end{figure}

The contributions to S and T parameters from the Littlest Higgs model with
T-parity are depicted in Fig.\ref{ST}. Constraints on S and T parameters from
electroweak precision measurements at 68\% and 99\% confidence level are also
shown. Solid lines are predictions of the model for $0 < R < 2$ and $f$ along
the center value of the U-branch (left figure) and L-branch (right
figure). For depicting contours of the constraints from the measurements, we
have used three experimental values following papers \cite{EWPO method}, $W$
boson mass ($m_W = 80.412 \pm 0.042$ GeV), weak mixing angle
($\sin^2\theta^{lept}_{eff} = 0.23153 \pm 0.00016$), and leptonic width of the
$Z$ boson ($\Gamma_l = 83.985\pm 0.086$ MeV) \cite{:2005em}. We have also used
values of the fine structure constant on the $Z$ pole ($\alpha^{-1}(m_Z) =
128.950\pm 0.048$) and the top quark mass ($m_t = 172.7\pm 2.9$ GeV)
\cite{unknown:2005cc}. The origin of the S-T plane is fixed by using the
reference Higgs mass, $m_h = 100$ GeV. In this figure, we can see that, even
for a large value of $f$, there is a parameter region consistent with the
electroweak precision measurements with appropriate choice of $R$. In the case
of large $f$, the Higgs boson becomes heavy and gives large contributions to
the S and T parameters. These contributions, however, can be canceled by those
from the top loops ($t$ and $T_+$) due to the large mixing angle ($R > 1$).

In Fig.\ref{STII}, constraints for the model parameters, $R$ and $f$, are shown
at 68\% and 99\% confidence level. At each point in these figures, the Higgs
mass is determined to satisfy the WMAP constraint on the U-branch (left figure)
and L-branch (right figure). Possible range of $R$ can be determined by
requiring $\lambda_1$ and $\lambda_2$ couplings to be within the perturbative
range, which results in $0.2 \lsim R \lsim 5$. As seen in these figures, the
entire allowed region can be consistent with the electroweak precision
measurements.


\begin{thebibliography}{99}

\bibitem{Barbieri}
R.~Barbieri and A.~Strumia,
Phys.\ Lett.\ B {\bf 433} (1998) 63;
R.~Barbieri and A.~Strumia,
arXiv:hep-ph/0007265.

\bibitem{Arkani-Hamed:2001nc}
N.~Arkani-Hamed, A.~G.~Cohen and H.~Georgi,
Phys.\ Lett.\ B {\bf 513} (2001) 232;
N.~Arkani-Hamed, A.~G.~Cohen, E.~Katz, A.~E.~Nelson, T.~Gregoire and
J.~G.~Wacker,
JHEP {\bf 0208} (2002) 021.

\bibitem{Arkani-Hamed:2002qy}
N.~Arkani-Hamed, A.~G.~Cohen, E.~Katz and A.~E.~Nelson,
JHEP {\bf 0207} (2002) 034.

\bibitem{difficulty}
C.~Csaki, J.~Hubisz, G.~D.~Kribs, P.~Meade and J.~Terning,
Phys.\ Rev.\ D {\bf 67} (2003) 115002;
J.~L.~Hewett, F.~J.~Petriello and T.~G.~Rizzo,
JHEP {\bf 0310} (2003) 062;
C.~Csaki, J.~Hubisz, G.~D.~Kribs, P.~Meade and J.~Terning,
Phys.\ Rev.\ D {\bf 68} (2003) 035009;
T.~Gregoire, D.~R.~Smith and J.~G.~Wacker,
Phys.\ Rev.\ D {\bf 69} (2004) 115008;
M.~C.~Chen and S.~Dawson,
Phys.\ Rev.\ D {\bf 70} (2004) 015003;
Z.~Han and W.~Skiba,
Phys.\ Rev.\ D {\bf 72} (2005) 035005;
W.~Kilian and J.~Reuter,
Phys.\ Rev.\ D {\bf 70} (2004) 015004.

\bibitem{Cheng:2003ju}
H.~C.~Cheng and I.~Low,
JHEP {\bf 0309} (2003) 051.

\bibitem{Cheng:2004yc}
H.~C.~Cheng and I.~Low,
JHEP {\bf 0408} (2004) 061.

\bibitem{Low:2004xc}
I.~Low,
JHEP {\bf 0410} (2004) 067.

\bibitem{Spergel:2003cb}
D.~N.~Spergel {\it et al.}  [WMAP Collaboration],
Astrophys.\ J.\ Suppl.\  {\bf 148} (2003) 175;
C.~L.~Bennett {\it et al.},
Astrophys.\ J.\ Suppl.\  {\bf 148} (2003) 1.

\bibitem{Jungman:1995df}
For reviews,\\
G.~Jungman, M.~Kamionkowski and K.~Griest,
Phys.\ Rept.\  {\bf 267} (1996) 195;
L.~Bergstrom,
Rept.\ Prog.\ Phys.\  {\bf 63}, (2000) 793;
G.~Bertone, D.~Hooper and J.~Silk,
Phys.\ Rept.\  {\bf 405} (2005) 279;
C.~Munoz,
Int.\ J.\ Mod.\ Phys.\ A {\bf 19} (2004) 3093.

\bibitem{Primack:2002th}
J.~R.~Primack,
Nucl.\ Phys.\ Proc.\ Suppl.\  {\bf 124} (2003) 3.

\bibitem{Hubisz:2004ft}
J.~Hubisz and P.~Meade,
Phys.\ Rev.\ D {\bf 71} (2005) 035016,
(For the correct paramter region consistent with the WMAP observation, 
see the figure in the revised vergion, hep-ph/0411264v3).

\bibitem{Hubisz:2005tx}
J.~Hubisz, P.~Meade, A.~Noble and M.~Perelstein,
JHEP {\bf 0601} (2006) 135.

\bibitem{positrons1}
S.~Rudaz and F.~W.~Stecker,
Astrophys.\ J.\  {\bf 325} (1988) 16;
J.~R.~Ellis, R.~A.~Flores, K.~Freese, S.~Ritz, D.~Seckel and J.~Silk,
Phys.\ Lett.\ B {\bf 214} (1988) 403;
A.~J.~Tylka,
Phys.\ Rev.\ Lett., {\bf 63} (1989) 840;
M.~S.~Turner and F.~Wilczek,
Phys.\ Rev.\ D, {\bf 42} (1990) 1001;
A.~J.~Tylka,
Phys.\ Rev.\ D {\bf 43} (1991) 1774.

\bibitem{positrons2}
E.~A.~Baltz, J.~Edsjo, K.~Freese and P.~Gondolo,
arXiv:astro-ph/0211239;
G.~L.~Kane, L.~T.~Wang and T.~T.~Wang,
Phys.\ Lett.\ B {\bf 536} (2002) 263;
W.~de Boer, C.~Sander, M.~Horn and D.~Kazakov,
Nucl.\ Phys.\ Proc.\ Suppl.\  {\bf 113} (2002) 221;
G.~L.~Kane, L.~T.~Wang and J.~D.~Wells,
Phys.\ Rev.\ D {\bf 65} (2002) 057701;
E.~A.~Baltz, J.~Edsjo, K.~Freese and P.~Gondolo,
Phys.\ Rev.\ D {\bf 65} (2002) 063511;
D.~Hooper and G.~D.~Kribs,
Phys.\ Rev.\ D {\bf 70} (2004) 115004;
S.~Profumo and P.~Ullio,
JCAP {\bf 0407} (2004) 006.

\bibitem{posbaltz}
E.~A.~Baltz and J.~Edsjo,
Phys.\ Rev.\ D {\bf 59} (1999) 023511.

\bibitem{Hooper:2004bq}
D.~Hooper and J.~Silk,
Phys.\ Rev.\ D {\bf 71} (2005) 083503.

\bibitem{antiprotons}
J.~Silk and M.~Srednicki,
Phys.\ Rev.\ Lett.\  {\bf 53} (1984) 624;
S.~Rudaz and F.~W.~Stecker, 
Astrophys.\ J. 325 (1988) 16;
A.~Bottino, F.~Donato, N.~Fornengo and P.~Salati,
Phys.\ Rev.\ D {\bf 58} (1998) 123503;
L.~Bergstrom, J.~Edsjo and P.~Ullio,
arXiv:astro-ph/9906034;
F.~Donato, N.~Fornengo, D.~Maurin, P.~Salati and R. Taillet,
Phys.\ Rev.\ D {\bf 69} (2004) 063501.

\bibitem{pamela}
M.~Circella  [PAMELA Collaboration],
Nucl.\ Instrum.\ Meth.\ A {\bf 518} (2004) 153;
P.~Spillantini,
Nucl.\ Instrum.\ Meth.\ B {\bf 214} (2004) 116.

\bibitem{Barao:2004ik}
F.~Barao  [AMS-02 Collaboration],
Nucl.\ Instrum.\ Meth.\ A {\bf 535} (2004) 134.

\bibitem{littlest_review}
M.~Schmaltz and D.~Tucker-Smith,
arXiv:hep-ph/0502182;
M.~Perelstein,
arXiv:hep-ph/0512128.

\bibitem{littlest}
G.~Burdman, M.~Perelstein and A.~Pierce,
Phys.\ Rev.\ Lett.\  {\bf 90} (2003) 241802
[Erratum-ibid.\  {\bf 92} (2004) 049903];
T.~Han, H.~E.~Logan, B.~McElrath and L.~T.~Wang,
Phys.\ Rev.\ D {\bf 67} (2003) 095004;
M.~Perelstein, M.~E.~Peskin and A.~Pierce,
Phys.\ Rev.\ D {\bf 69} (2004) 075002.

\bibitem{Birkedal:2006fz}
A.~Birkedal, A.~Noble, M.~Perelstein and A.~Spray,
arXiv:hep-ph/0603077.

\bibitem{Kolb:1990vq}
E.~W.~Kolb and M.~S.~Turner,
{\it The Early Universe},
(Addison-Wesley, Reading, MA, 1990).

\bibitem{profile}
J.~F.~Navarro, C.~S.~Frenk and S.~D.~M.~White,
Astrophys.\ J.\  {\bf 462} (1996) 563;
J.~F.~Navarro, C.~S.~Frenk and S.~D.~M.~White,
Astrophys.\ J.\  {\bf 490} (1997) 493;
B.~Moore, S.~Ghigna, F.~Governato, G.~Lake, T.~Quinn, J.~Stadel and P.~Tozzi,
Astrophys.\ J.\  {\bf 524} (1999) L19;
J.~Diemand, B.~Moore and J.~Stadel,
Mon.\ Not.\ Roy.\ Astron.\ Soc.\  {\bf 353} (2004) 624.

\bibitem{Corcella:2000bw}
G.~Corcella {\it et al.},
interfering gluons (including supersymmetric processes),''
JHEP {\bf 0101} (2001) 010.

\bibitem{Hisano:2005ec}
J.~Hisano, S.~Matsumoto, O.~Saito and M.~Senami,
arXiv:hep-ph/0511118.

\bibitem{Boost Factor}
J.~Silk and A.~Stebbins,
Astrophys.\ J.\  {\bf 411} (1993) 439;
L.~Bergstrom, J.~Edsjo and P.~Gondolo,
Phys.\ Rev.\ D {\bf 59} (1999) 043506.

\bibitem{Berezinsky:2003vn}
V.~Berezinsky, V.~Dokuchaev and Y.~Eroshenko,
Phys.\ Rev.\ D {\bf 68} (2003) 103003.

\bibitem{BtoC}
A.~W.~Strong and I.~V.~Moskalenko,
Astrophys.\ J.\  {\bf 509} (1998) 212;
I.~V.~Moskalenko and A.~W.~Strong,
Phys.\ Rev.\ D {\bf 60} (1999) 063003;
D.~Maurin, F.~Donato, R.~Taillet and P.~Salati,
Astrophys.\ J.\  {\bf 555} (2001) 585;
D.~Maurin, R.~Taillet and F.~Donato,
Astron.\ Astrophys.\  {\bf 394} (2002) 1039;
I.~V.~Moskalenko, A.~W.~Strong, S.~G.~Mashnik and J.~F.~Ormes,
Astrophys.\ J.\  {\bf 586} (2003) 1050;
D.~Hooper, J.~E.~Taylor and J.~Silk,
Phys.\ Rev.\ D {\bf 69} (2004) 103509.

\bibitem{Longair:1994wu}
 M.~S.~Longair,
{\it High-energy astrophysics. Vol. 2}
Cambridge University Press, New York, 1994.

\bibitem{Barrau:2001ev}
A.~Barrau, G.~Boudoul, F.~Donato, D.~Maurin, P.~Salati and R.~Taillet,
Astron.\ Astrophys.\  {\bf 388} (2002) 676.

\bibitem{solarmod1}
L.~J.~Gleeson and W.~I.~Axford,
Astrophys.\ J.\  {\bf 149} (1967) L115.

\bibitem{Moskalenko:1997gh}
I.~V.~Moskalenko and A.~W.~Strong,
Astrophys.\ J.\  {\bf 493} (1998) 694.

\bibitem{HEAT}
S.~W.~Barwick {\it et al.}  [HEAT Collaboration],
Phys.\ Rev.\ Lett.\  {\bf 75} (1995) 390.
S.~W.~Barwick {\it et al.}  [HEAT Collaboration],
Astrophys.\ J.\  {\bf 482} (1997) L191;

\bibitem{Drollinger:2001bc}
V.~Drollinger and A.~Sopczak,
Eur.\ Phys.\ J.\ directC {\bf 3} (2001) N1.

\bibitem{Meyer:2004ha}
N.~Meyer and K.~Desch,
Eur.\ Phys.\ J.\ C {\bf 35} (2004) 171.

\bibitem{Alsharoa:2002wu}
M.~M.~Alsharoa {\it et al.}
[Muon Collider/Neutrino Factory Collaboration],
Phys.\ Rev.\ ST Accel.\ Beams {\bf 6} (2003) 081001.

\bibitem{Peskin:1991sw}
M.~E.~Peskin and T.~Takeuchi,
Phys.\ Rev.\ D {\bf 46} (1992) 381.

\bibitem{Feynman formula}
See, for exapmle, p.12,
Yuri Makeenko, ``{\it Method of Contemporary Gauge Theory}''
(Cambridge university press).

\bibitem{EWPO method}
W.~J.~Marciano,
arXiv:hep-ph/0003181;
M.~Perelstein, M.~E.~Peskin and A.~Pierce,
Phys.\ Rev.\ D {\bf 69} (2004) 075002.

\bibitem{:2005em}
[ALEPH Collaboration],
Phys.\ Rept.\  {\bf 427} (2006) 257.

\bibitem{unknown:2005cc}
[CDF Collaboration],
arXiv:hep-ex/0507091.

\end{thebibliography}
\end{document}